\documentclass[a4paper]{spie}

\usepackage[a4paper,
            left=1.93cm,
            right=1.93cm,
            top=2.54cm,
            bottom=4.94cm]{geometry}

\usepackage{graphicx}
\usepackage{amsmath,amsfonts,amssymb}
\usepackage{booktabs}
\usepackage{multirow}
\usepackage{caption, subcaption}
\usepackage[colorlinks=true, allcolors=blue]{hyperref}
\usepackage{float}

\title{Automated multi-stressor optical fiber focal ratio degradation and throughput characterization system for future massive multiplex spectroscopic survey}

\author[a,b,*]{Daksh Singh}
\author[a,c]{Ting S. Li}
\author[a]{Sarik Jeram}
\author[c]{Shaojie Chen}

\affil[a]{David A. Dunlap Department of Astronomy and Astrophysics, University of Toronto, 50 St. George Street, Toronto, Ontario, M5S 3H4, Canada }
\affil[b]{Department of Physics, University of Toronto, 50 St. George Street, Toronto, Ontario,  M5S 3H4, Canada}
\affil[c]{Dunlap Institute for Astronomy and Astrophysics, University of Toronto, 50 St. George Street, Toronto, Ontario, M5S 3H4, Canada}

\authorinfo{Further author information: \\
Daksh Singh: daksh.singh@mail.utoronto.ca}

\begin{document}

\maketitle

\begin{abstract}
Future spectroscopic survey facilities such as Spec-S5 and the Wide-Field Spectroscopic Telescope (WST) will contain more than 25,000 optical fibers. Characterizing the focal ratio degradation (FRD) and throughput of these fibers is essential for spectroscopic calibration, sky subtraction, and overall instrument performance, but manual testing of such large fiber populations is prohibitively time-consuming. In this work, we present an automated multi-stressor optical fiber characterization platform capable of measuring both FRD and relative throughput while systematically varying optical and mechanical stress conditions. The system enables characterization as a function of input angle, wavelength, bend radius, twist angle, and externally applied load, while also allowing combinations of stressors to be investigated simultaneously. A custom software framework automates hardware control, data acquisition, and data analysis, enabling large characterization campaigns to be performed with minimal user intervention. Validation measurements performed on a multimode optical fiber demonstrate the ability of the platform to resolve both pronounced and subtle stress-dependent variations in FRD and relative throughput. The results show the expected dependence of FRD on input angle and bend radius, reveal comparatively weak sensitivity to externally applied compressive loading under the investigated conditions, and identify a reproducible periodic dependence on fiber twist angle. In addition, relative throughput measurements indicate that the reduction in relative throughput with increasing input angle cannot be explained solely by a simple projected-area model. The automated workflow and scalability of the system provide a versatile framework for the large-scale characterization of optical fibers required by next-generation spectroscopic survey facilities.
\end{abstract}

\section{INTRODUCTION}
Modern astronomical spectroscopy increasingly depends on highly multiplexed fiber-fed instrumentation. Next-generation facilities, such as the Spectroscopic Stage-5 Experiment (Spec-S5) and the Wide-Field Spectroscopic Telescope (WST), are expected to use more than 25,000 optical fibers to transmit light from telescope focal planes to spectrographs \cite{spec5,wst}. Maintaining high optical performance across these large fiber populations is essential for maximizing scientific output and ensuring stable spectroscopic calibration \cite{Monty2018}.

Focal ratio degradation (FRD) and throughput are two primary performance metrics for optical fibers used in astronomical instrumentation. Throughput quantifies the fraction of incident optical power transmitted through a fiber, while FRD describes the increase in divergence of the output beam relative to the injected beam. FRD arises from modal redistribution caused by microscopic imperfections, stress-induced perturbations, bending, twisting, compression, and manufacturing or termination defects \cite{Gloge1972,Belland2020,Wang2023}. Excessive FRD reduces spectrograph efficiency, alters the illumination of spectrograph optics, changes point-spread functions, and complicates calibration procedures. Losses in throughput similarly reduce signal-to-noise ratio and observing efficiency. Consequently, comprehensive characterization of fiber performance is necessary before installation in large-scale spectroscopic facilities \cite{Belland2020,Wang2023,Monty2018}.

The importance of FRD in astronomical instrumentation has been recognized for several decades. Early studies of fiber optical waveguides for astronomy demonstrated that fibers can significantly alter the angular distribution of transmitted light, motivating the need to quantify focal-ratio preservation as part of instrument design \cite{Barden1981,Ramsey1988}. Subsequent work developed practical methods for measuring and modeling FRD in astronomical fibers. Carrasco and Parry introduced a widely used method for characterizing FRD using the output angular distribution of light from a fiber \cite{Carrasco1994}, while later studies refined these approaches by investigating the dependence of FRD on fiber length, termination, polishing, bonding, and end effects \cite{Poppett2010,AllingtonSmith2013,Eigenbrot2012}. These studies showed that FRD is not solely an intrinsic property of the bulk fiber, but can also depend strongly on how the fiber is prepared, mounted, polished, bonded, routed, and integrated into an instrument.

More recent work has shifted from isolated fiber characterization toward instrument-specific testing and system-level performance validation. Studies of fibers and fiber assemblies for instruments such as VIRUS-P, DESI, Subaru/PFS, MSE, and Hector have examined how FRD and throughput are affected by wavelength, input focal ratio, fiber motion, bending, positioner operation, fiber bundles, and integral-field-unit fabrication \cite{Murphy2008,PoppettDESI2014,Santos2014,Monty2018,Belland2020,Wang2023}. These efforts reflect the current state of the field: FRD characterization is increasingly treated as a quality-assurance and risk-reduction problem for large fiber-fed spectroscopic facilities, rather than only as a laboratory measurement of individual fibers.

Despite this progress, there remains a need for scalable characterization systems that can measure both FRD and throughput while systematically varying multiple optical and mechanical stress conditions. Many existing test benches focus on a single fiber property, a single stressor, or an instrument-specific configuration. Future massively multiplexed surveys intensify this challenge because they require repeatable characterization of thousands to tens of thousands of fibers and fiber-positioner assemblies. For example, DESI uses 5000 fibers \cite{Poppett2024}, MSE test-bench development was motivated by the need to characterize thousands of fibers \cite{Monty2018}, and WST concepts require fiber-positioning systems with tens of thousands of robotic units \cite{wst}. These requirements motivate automated platforms capable of measuring fiber performance under realistic combinations of stress, alignment, wavelength, and coupling conditions.

In this work, we present an automated multi-stressor optical fiber characterization platform that measures both FRD and relative throughput while systematically varying input angle, bend radius, twist angle, externally applied load, and wavelength. By integrating multiple stressor subsystems within a single automated framework, the platform enables rapid, repeatable, and scalable characterization of optical fibers under conditions representative of astronomical instrumentation. This approach provides a practical framework for evaluating fiber performance in preparation for future large-scale fiber-fed spectroscopic survey facilities.

\section{THEORY}

\subsection{Focal Ratio Degradation}
\begin{figure}[ht]
\centering
\includegraphics[width=0.5\textwidth]{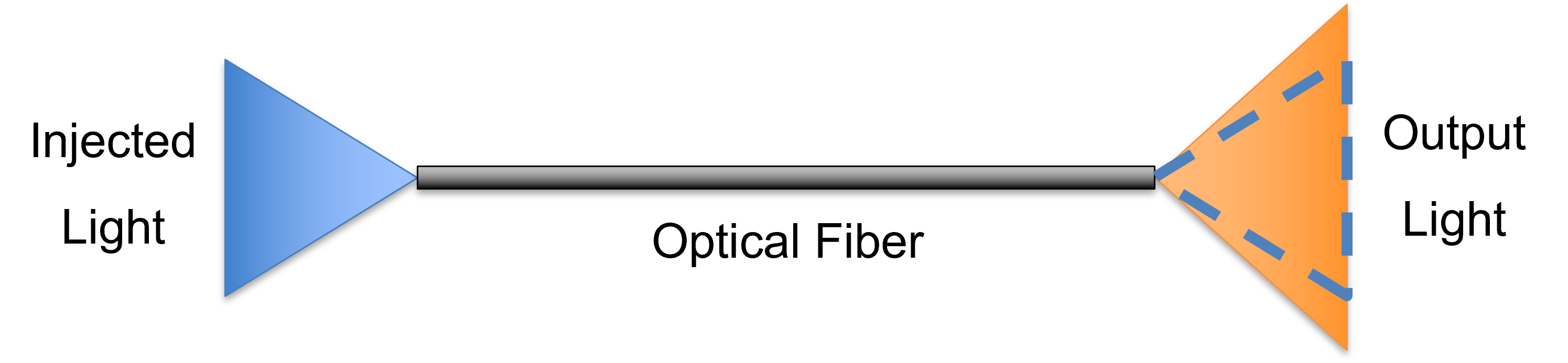}
\caption{Schematic illustration of focal ratio degradation (FRD) in an optical fiber. The blue cone represents the injected light cone, while the orange cone represents the output light cone after propagation through the fiber. The dashed blue cone indicates the expected output cone in the absence of FRD. The increased divergence of the orange cone illustrates the degradation of the focal ratio caused by scattering and mode coupling within the optical fiber.}
\label{fig:FRD_diagram}
\end{figure}

FRD occurs when scattering and mode-coupling mechanisms within an optical fiber redistribute optical power among propagating modes. These mechanisms can be caused by surface imperfections, microbending, bending, twisting, compression, and other stress-induced perturbations that alter light propagation within the fiber \cite{Belland2020,Wang2023}. As a result, the output beam exits the fiber with a larger angular spread than the injected beam, corresponding to a degradation of the focal ratio, as illustrated in Figure~\ref{fig:FRD_diagram}. FRD can be described theoretically using the modal diffusion model developed by Gloge \cite{Gloge1972} and later analyzed by Gambling et al. \cite{Gambling1975}.

A common method for quantifying FRD is the collimated beam (ring) test \cite{Carrasco1994,AllingtonSmith2013,Monty2018}, in which a collimated beam is injected into a fiber at an angle $\theta$ relative to the fiber axis. The far-field intensity distribution forms a ring whose radius corresponds to the injection angle and whose thickness is related to the degree of modal scattering.

Adapting the approach of Monty et al. \cite{Monty2018}, the input f/\# is related to the input angle by

\begin{align}
    f/\#_{in}=\frac{1}{2\tan|\theta|},
\label{eq:f_num_in}
\end{align}

\noindent and FRD may be quantified using the ratio of the ring thickness to the ring diameter:

\begin{equation}
    \text{FRD} [\%]=\frac{\text{Ring Thickness}}{D}\times100,
    \label{eq:FRD_equation}
\end{equation}

\noindent where $D$ is the diameter of the far-field ring and $\mathrm{Ring~Thickness}$ is the characteristic width of the ring. Since both quantities are measured from the same detector image, their ratio is dimensionless and can be calculated directly using pixel units. In this work, the ring thickness is determined from the full width at half maximum (FWHM) of the radial intensity profile generated from the far-field image. Specifically, once the FWHM has been found, the ring thickness is taken to be the corresponding $1/e^2$ diameter of a Gaussian profile:

\begin{equation}
     \text{Ring Thickness} =\text{$1/e^2$ Diameter} = \sqrt\frac{2}{\ln2}\times\text{FWHM}.
     \label{eq:beam_diameter}
 \end{equation}

\begin{figure}[ht]
    \centering
    \includegraphics[width=0.8\textwidth]{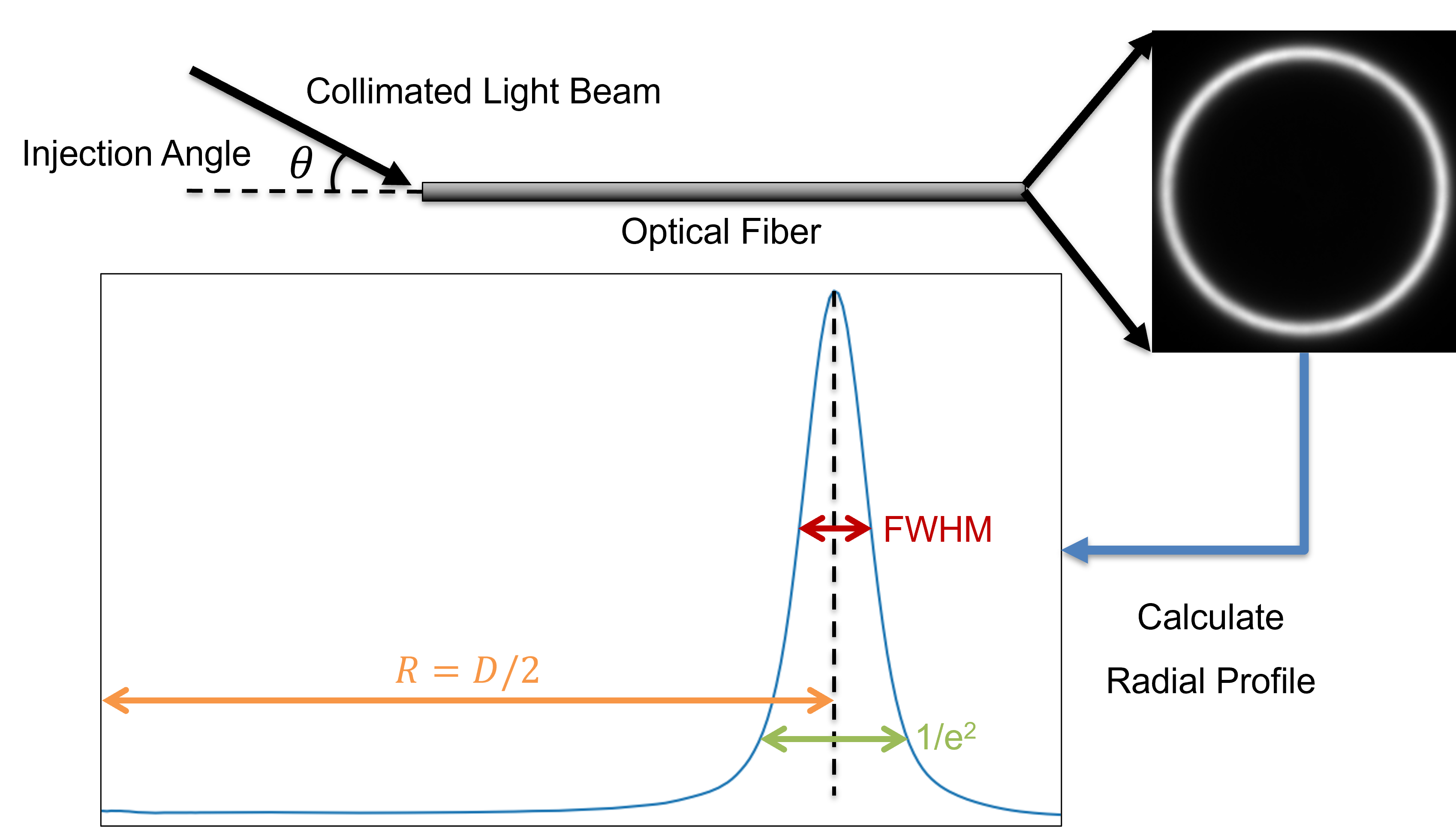}
    \caption{Overview of the procedure used to determine focal ratio degradation (FRD) from a far-field intensity pattern obtained using the collimated beam (ring) test.}
    \label{fig:FRD_measurement_overview}
\end{figure}

Figure~\ref{fig:FRD_measurement_overview} illustrates the collimated beam (ring) test used to measure FRD. A collimated light beam is injected into the optical fiber at an angle relative to the fiber axis, producing a ring-shaped far-field intensity pattern at the fiber output. A radial intensity profile is then generated from the far-field image, allowing the ring diameter and ring thickness to be determined. These quantities are subsequently used to calculate the FRD according to Equation~\ref{eq:FRD_equation}.

\subsection{Relative Throughput}

Throughput is generally defined as the fraction of optical power transmitted through an optical fiber and may be expressed as

\begin{equation}
    \text{Throughput} [\%]=\frac{P_{out}}{P_{in}}\times100,
    \label{eq:throughput_equation}
\end{equation}

\noindent where $P_{in}$ and $P_{out}$ are the optical powers measured at the fiber input and output, respectively. Throughput measurements complement FRD measurements by quantifying changes in transmitted optical power under different optical and mechanical stress conditions.

In the present implementation, the quantity reported using Equation~\ref{eq:throughput_equation} is treated as a relative throughput metric. This is because the input photodiode measures the optical power in the reference arm of the beam splitter rather than the optical power coupled directly into the fiber core. As a result, the reported values should be interpreted as relative measurements used to compare changes between configurations, rather than as absolute measurements of the fiber transmission efficiency.

The input and output optical powers are measured simultaneously. Therefore, temporal fluctuations in the light-source intensity affect both measurements and largely cancel when calculating the ratio in Equation~\ref{eq:throughput_equation}. This reduces the sensitivity of the relative throughput measurement to source-power variations during data acquisition.

\section{SYSTEM OVERVIEW}

\begin{figure}[ht]
\centering
\includegraphics[width=\textwidth]{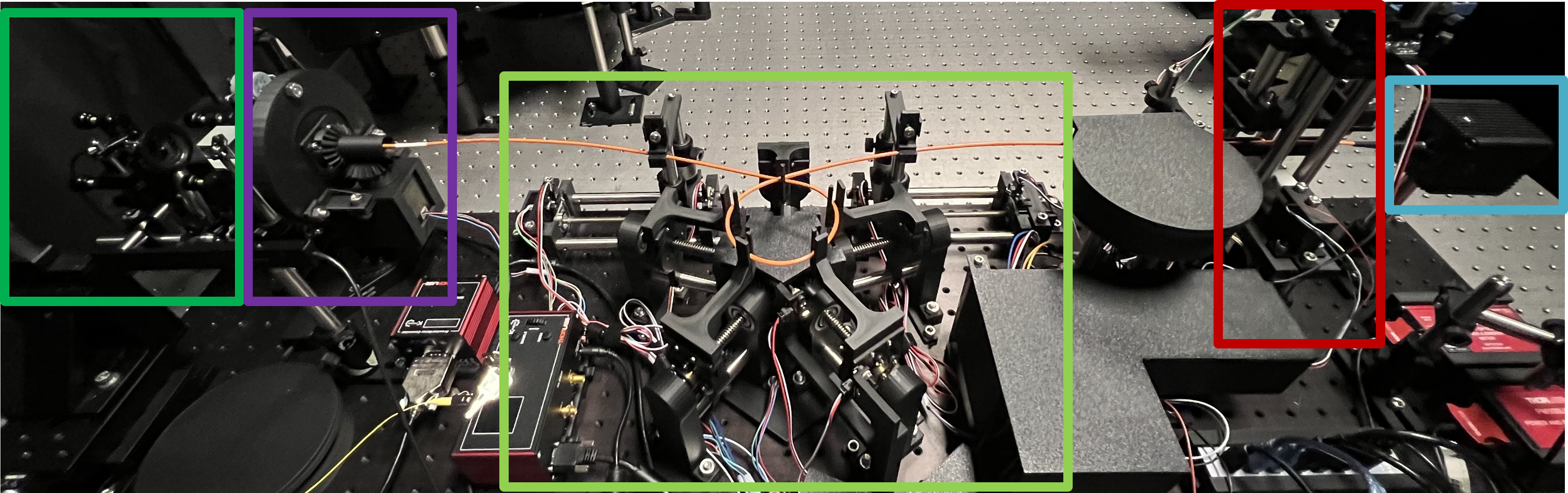}
\caption{Photograph of the fully automated FRD and relative throughput characterization system. The optical fiber under test is highlighted in orange. The input beam control, twist-angle control, bend-radius control, weight-application, and detection subsystems are highlighted in dark green, purple, light green, red, and blue, respectively. These color assignments are used consistently throughout the manuscript to identify the corresponding subsystem photographs.}
\label{fig:full_system}
\end{figure}

This section describes the major components of the automated multi-stressor optical fiber characterization platform and their functions. An image of the complete system is shown in Figure~\ref{fig:full_system}. The system consists of subsystems for input beam control, bend-radius control, twist-angle control, weight-application, far-field intensity pattern acquisition, and relative throughput measurement, all of which are coordinated through a custom software framework.

\subsection{Hardware}

\subsubsection{Wavelength Selection}
\label{sec:wl_selection}
The system was illuminated using Thorlabs EP470S04, EP590S04, and EP700S04 visible pigtailed LEDs. The specifications of these light sources are summarized in Table~\ref{tab:light_sources}. To change the illumination wavelength, the installed pigtailed LED must be manually exchanged for the desired source.

\begin{table}[ht]
\caption{Specifications of the light sources used in this work. Values were obtained from the manufacturer-provided datasheets for the individual LEDs \cite{Thorlabs_EP470S04_Datasheet, Thorlabs_EP590S04_Datasheet, Thorlabs_EP700S04_Datasheet}.} 
\label{tab:light_sources}
\begin{center}       
\begin{tabular}{|l|l|l|l|} 
\hline
\rule[-1ex]{0pt}{3.5ex}  Thorlabs Part Number & Center Wavelength & Bandwidth & Output Power  \\
\hline
\rule[-1ex]{0pt}{3.5ex}  EP470S04 & 484.9 nm & 25 nm & 18 mW  \\
\hline
\rule[-1ex]{0pt}{3.5ex}  EP590S04 & 594.2 nm & 14 nm & 3.5 mW \\
\hline
\rule[-1ex]{0pt}{3.5ex}  EP700S04 & 697.6 nm & 18 nm & 2 mW \\
\hline 
\end{tabular}
\end{center}
\end{table}

\subsubsection{Input Beam Control Subsystem}
\label{sec:input}
\begin{figure}[htb]
\centering
\includegraphics[width=0.8\textwidth]{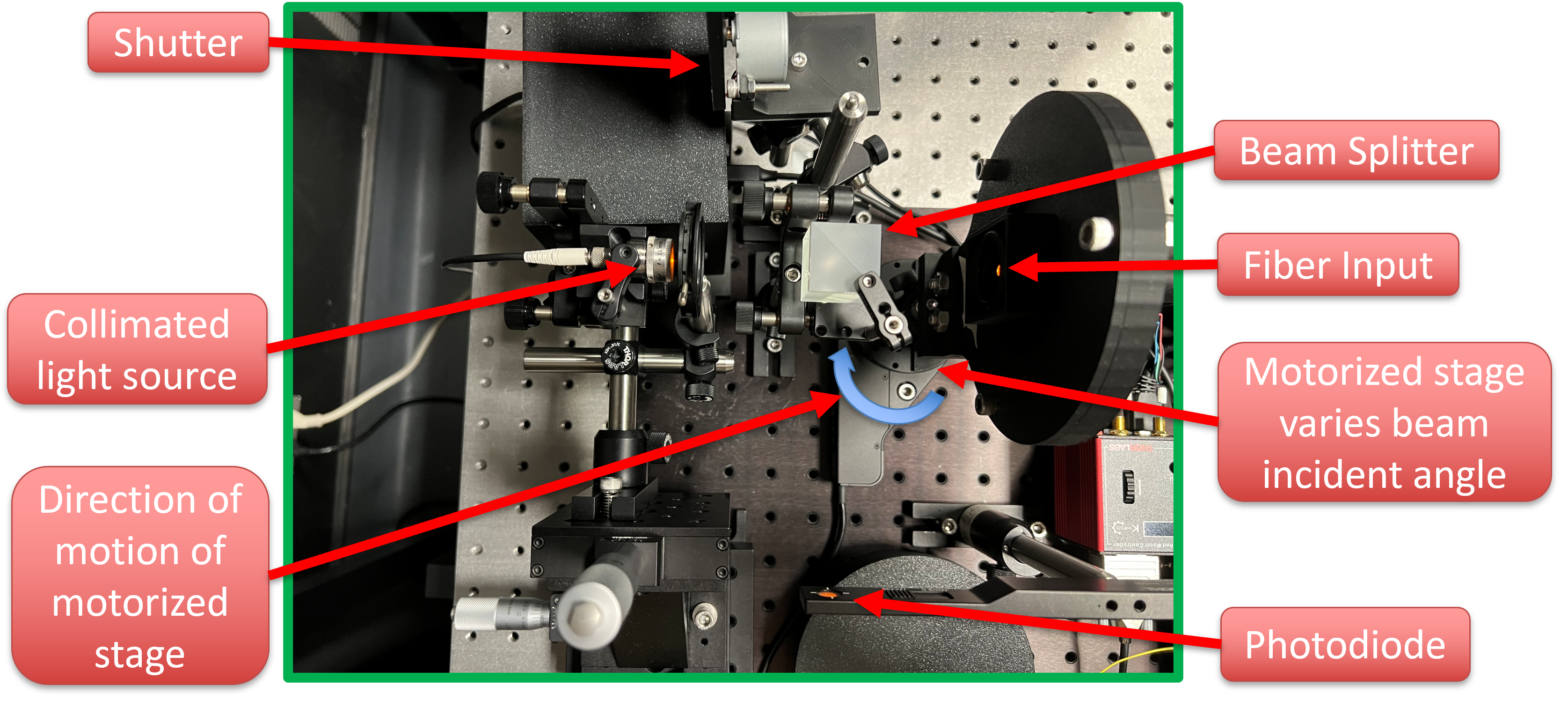}
\caption{Photograph of the input beam control subsystem used to adjust the illumination conditions during FRD and relative throughput measurements. The optical fiber under test is shown in orange. The blue arrow indicates the direction of motion of the motorized rotation stage used to vary the incident angle of the input beam.}
\label{fig:input_subsystem}
\end{figure}

Figure~\ref{fig:input_subsystem} shows the input beam control subsystem. The light sources described in Section~\ref{sec:wl_selection} are coupled to a Thorlabs CFCS11-A fiber collimator to produce a collimated beam that is injected into the optical fiber under test. An iris is positioned after the collimator to isolate a narrow beam that illuminates the fiber input. A servo-actuated shutter can also be inserted into the beam path to block incident light from entering the optical fiber, enabling automated acquisition of dark frames for background-light subtraction.

A nominal 50:50 beam splitter directs a portion of the incident light to a Thorlabs S130VC photodiode power sensor, while the remaining light is directed toward the optical fiber under test. Prior to use, the photodiode is calibrated to the wavelength of the light source and is used to monitor the input optical power for relative throughput calculations. This arrangement enables simultaneous measurements of input and output optical power, reducing the effect of source-intensity fluctuations on relative throughput measurements.

The incident angle of the input beam is controlled using a Thorlabs PRMTZ8 motorized continuous-rotation stage, implemented following the approach described by Monty et al.~\cite{Monty2018}. The angular position of the stage is restricted to the range $-15.00^{\circ}\pm0.05^{\circ}$ to $15.00^{\circ}\pm0.05^{\circ}$, excluding $0^{\circ}$, where the input f/\# is undefined according to Equation~\ref{eq:f_num_in}.

\subsubsection{Twist Angle Control Subsystem}
\begin{figure}[htb]
\centering

\begin{subfigure}{0.2\textwidth}
    \includegraphics[width=\textwidth]{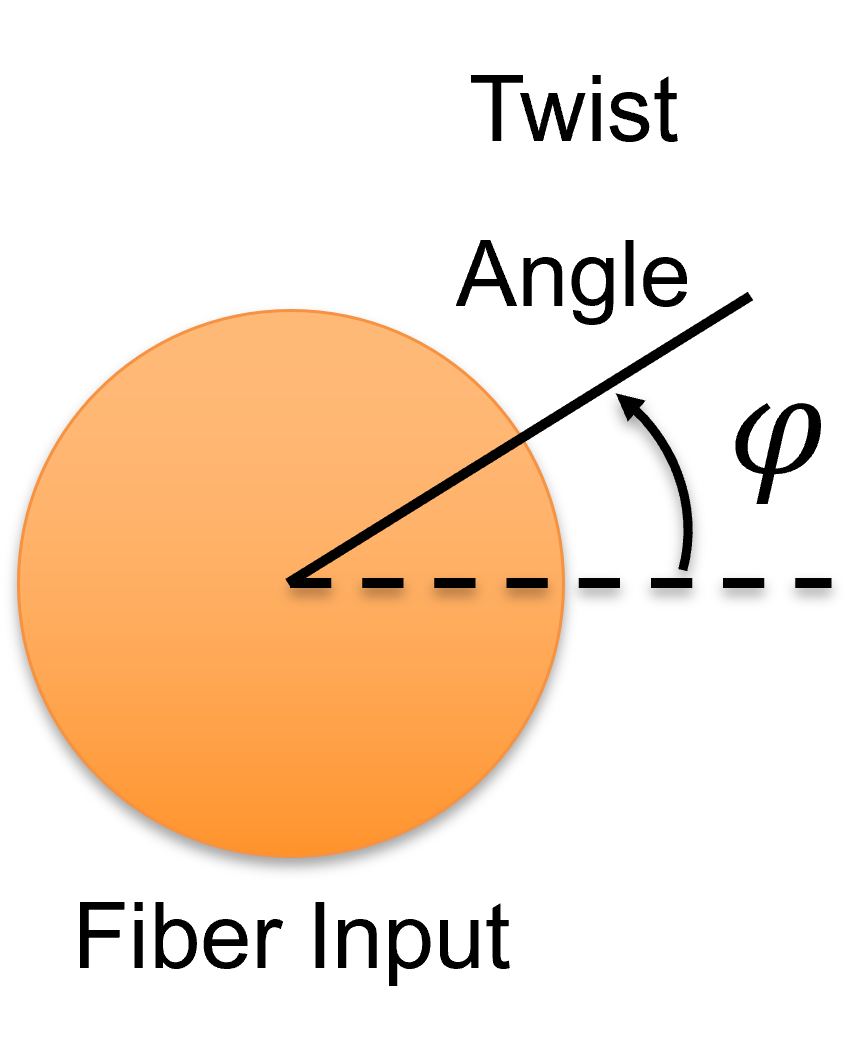}
    \caption{}
    \label{fig:twist_diagram}
\end{subfigure}
~
\begin{subfigure}{0.35\textwidth}
    \includegraphics[width=\textwidth]{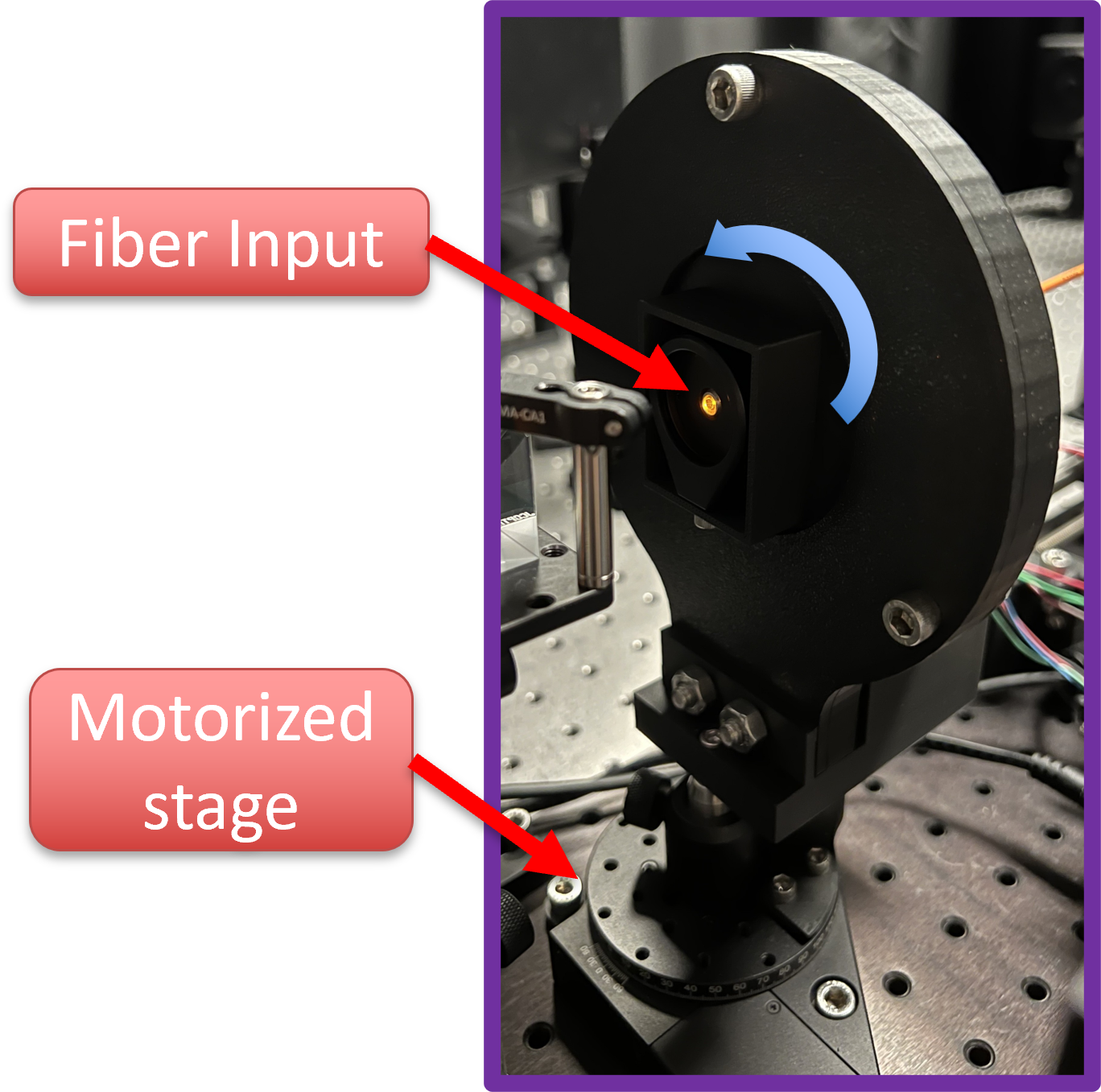}
    \caption{}
    \label{fig:twist_front}
\end{subfigure}
~
\begin{subfigure}{0.231\textwidth}
    \includegraphics[width=\textwidth]{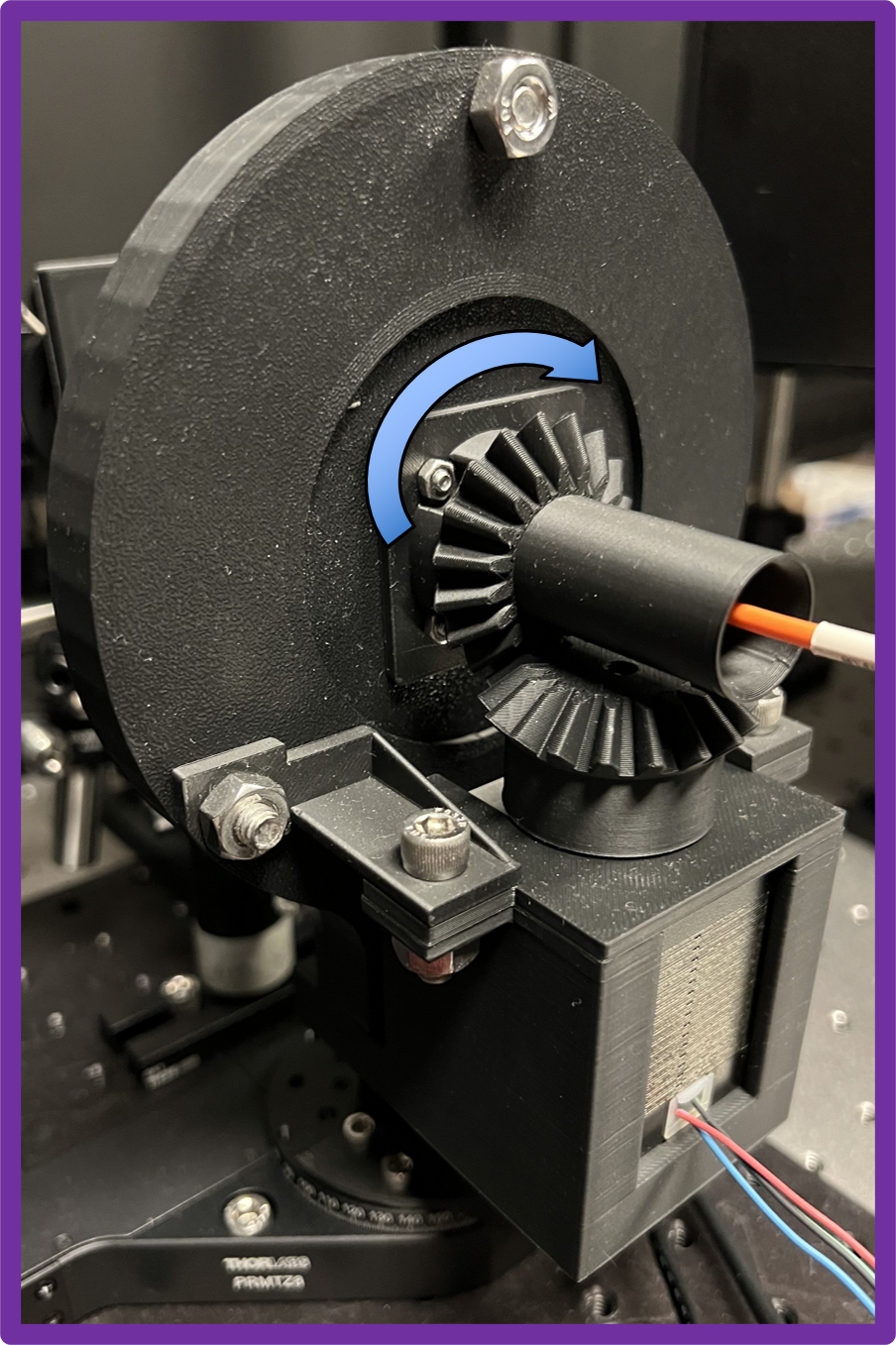}
    \caption{}
    \label{fig:twist_back}
\end{subfigure}

\caption{Twist-angle control subsystem used to apply controlled torsional stress to the optical fiber during FRD and relative throughput measurements. (a) Definition of the twist angle as viewed from the fiber input. (b) Front view of the twist-angle control subsystem. (c) Rear view of the twist-angle control subsystem. The optical fiber under test is highlighted in orange, and the blue arrow indicates the direction of rotation used to vary the twist angle.}
\label{fig:twist_subsystem}
\end{figure}

Figures~\ref{fig:twist_front} and~\ref{fig:twist_back} show the front and rear views of the twist-angle control subsystem, respectively. As illustrated in Figure~\ref{fig:twist_diagram}, the subsystem rotates the input end of the optical fiber about its optical axis while maintaining the input coupling geometry. This enables controlled investigation of the effects of torsional stress on FRD and relative throughput. Rotation is driven by a stepper motor, providing repeatable angular positioning. Although the subsystem is mechanically capable of larger rotations, the operating range is limited to $-360.00^{\circ}\pm0.09^{\circ}$ to $360.00^{\circ}\pm0.09^{\circ}$ to prevent excessive twisting and potential damage to the optical fiber.

\subsubsection{Bend-Radius Control Subsystem}
\begin{figure}[htb]
\centering

\begin{subfigure}{0.3\textwidth}
    \includegraphics[width=\textwidth]{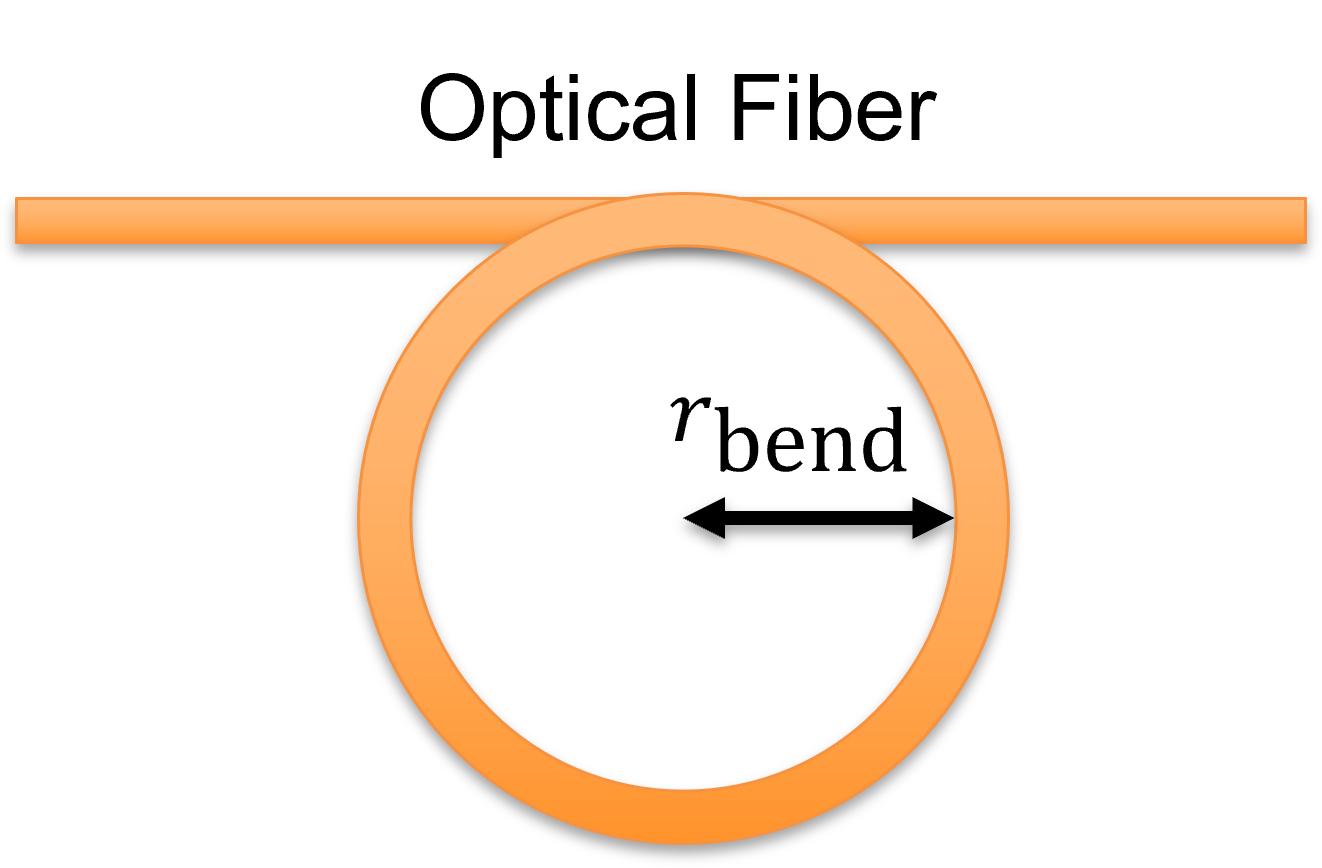}
    \caption{}
    \label{fig:bend_diagram}
\end{subfigure}
~
\begin{subfigure}{0.39\textwidth}
    \includegraphics[width=\textwidth]{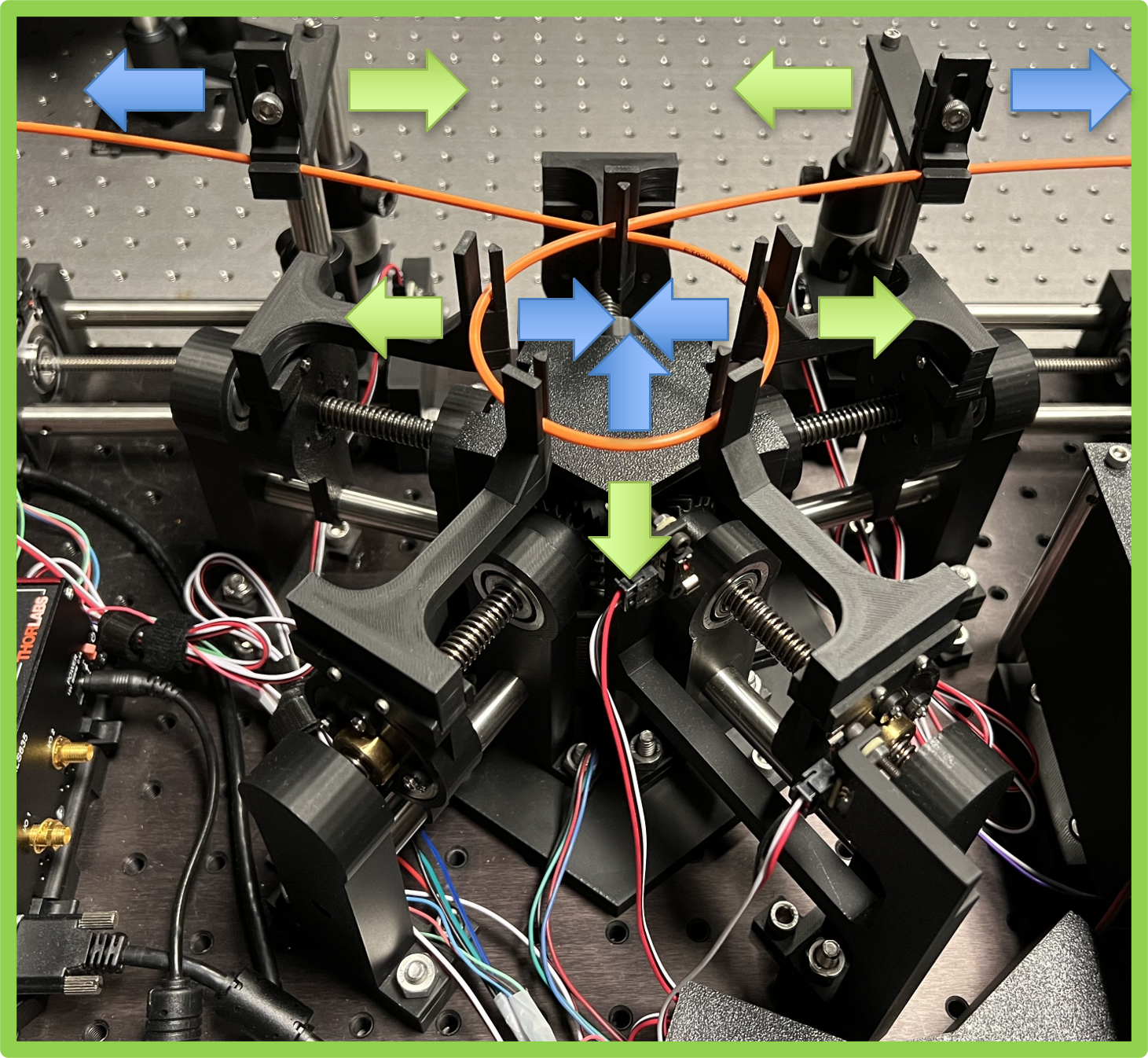}
    \caption{}
    \label{fig:bend_pic}
\end{subfigure}

\caption{Bend-radius control subsystem used to apply controlled bending to the optical fiber during FRD and relative throughput measurements. (a) Definition of the bend radius. (b) Bend-radius control subsystem. The optical fiber under test is highlighted in orange. The green arrows indicate the direction of motion of the fiber segments as the bend radius increases, while the blue arrows indicate the direction of motion as the bend radius decreases.}
\label{fig:bend_radius_subsystem}
\end{figure}

Figure~\ref{fig:bend_pic} shows the bend-radius control subsystem, which enables characterization of the effects of controlled bending on FRD and relative throughput. A dedicated mechanical assembly controls the bend radius of the optical fiber, while motorized actuation provides repeatable adjustment of the radius of curvature with minimal unintended perturbations. In this work, the bend radius is defined as the distance from the center of the fiber loop to the inner curvature of the fiber, as illustrated in Figure~\ref{fig:bend_diagram}. The subsystem provides a bend-radius range of $5\pm1$~mm to $36\pm1$~mm.

\subsubsection{Weight-Application Subsystem}
\begin{figure}[htb]
\centering

\begin{subfigure}{0.35\textwidth}
    \includegraphics[width=\textwidth]{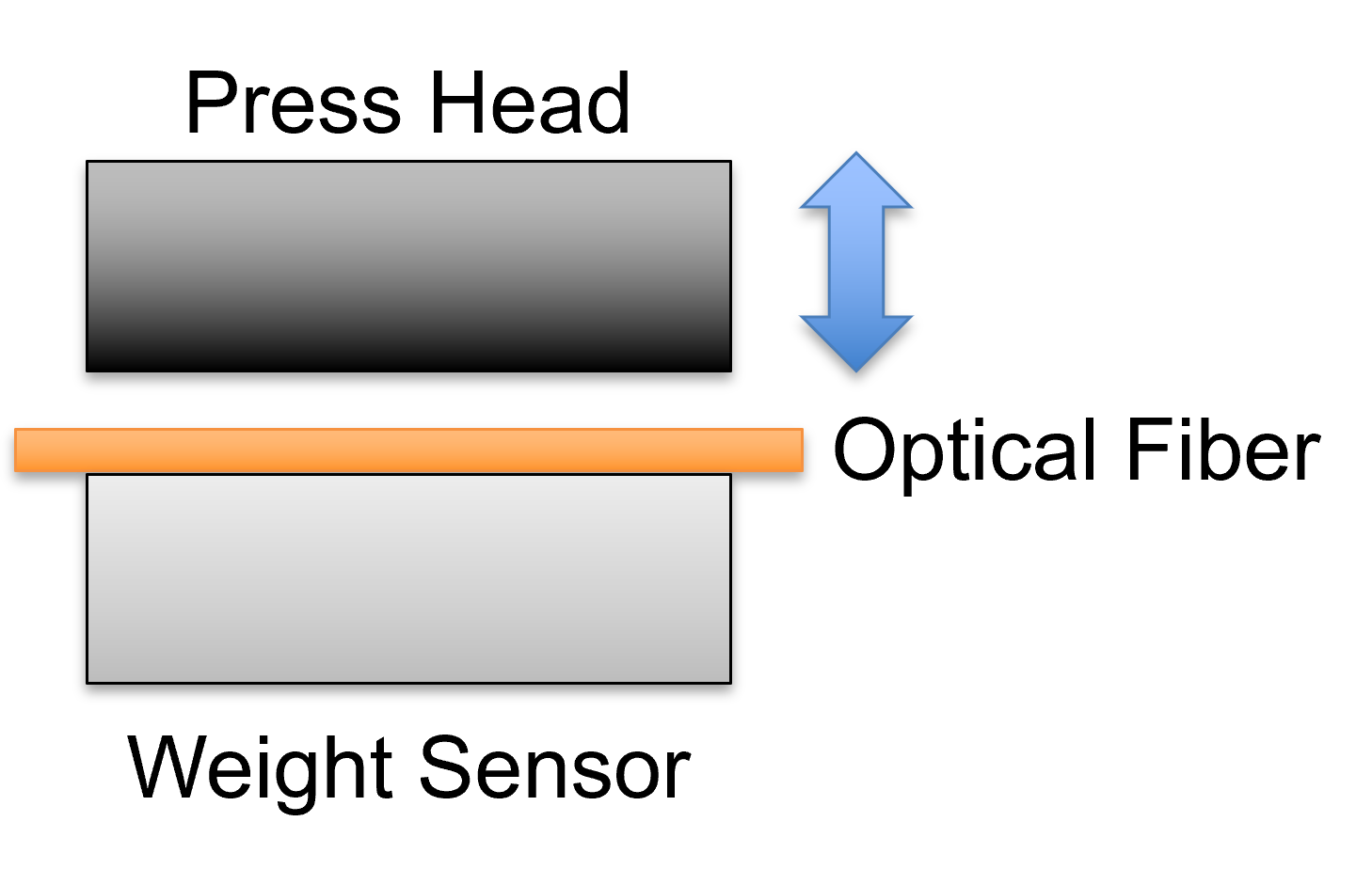}
    \caption{}
    \label{fig:load_diagram}
\end{subfigure}
~
\begin{subfigure}{0.23\textwidth}
    \includegraphics[width=\textwidth]{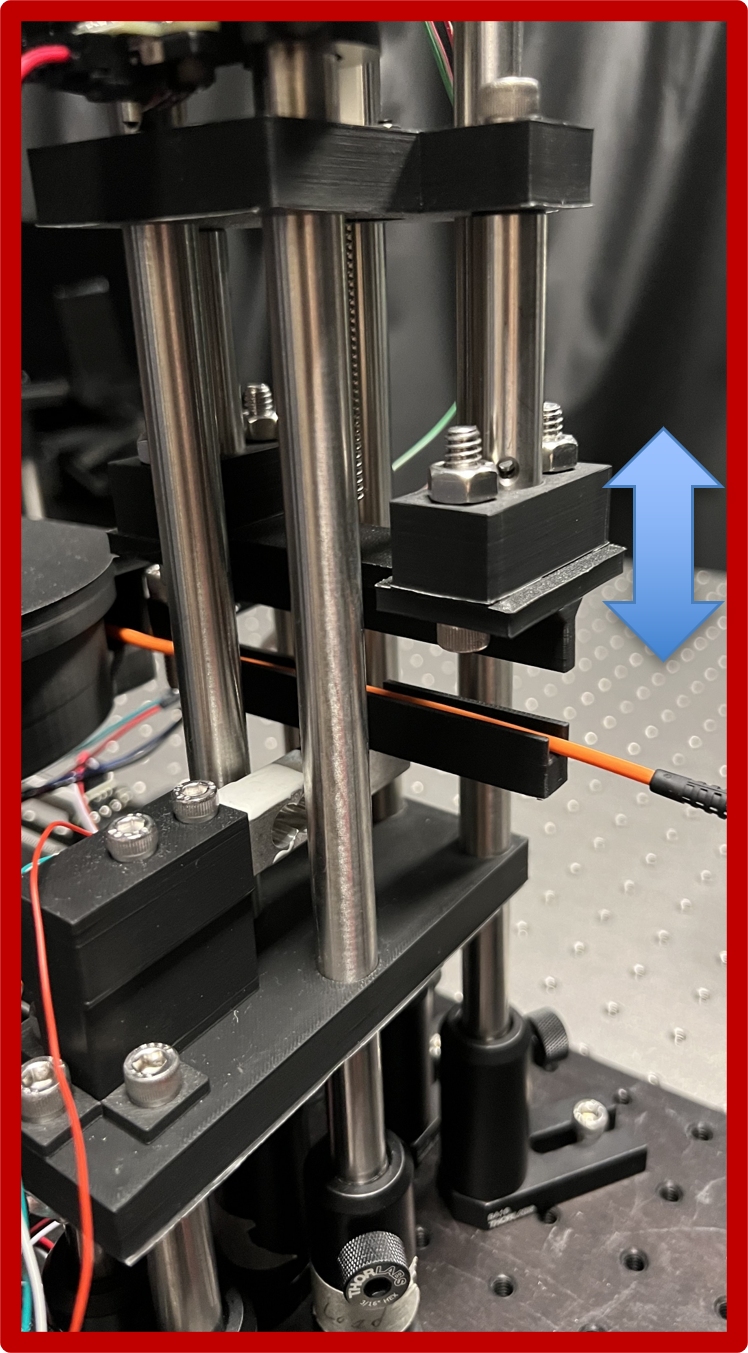}
    \caption{}
    \label{fig:load_pic}
\end{subfigure}
\caption{Weight-application subsystem used to apply controlled compressive loads to the optical fiber during FRD and relative throughput measurements. (a) Schematic showing the relative locations of the press head, optical fiber, and weight sensor. (b) Photograph of the weight-application subsystem. The optical fiber under test is highlighted in orange, and the blue arrow indicates the direction of motion of the press head used to apply the compressive load.}
\label{fig:load_subsystem}
\end{figure}

Figure~\ref{fig:load_pic} shows the weight-application subsystem, which enables investigation of the effects of controlled compressive loading on FRD and relative throughput. A stepper motor rotates a lead screw to translate a press head vertically. When the press head contacts the optical fiber, it applies a controlled compressive load over a length of $100\pm1$~mm. A weight sensor continuously monitors the applied load to ensure repeatable force measurements throughout testing. The motion of the press head, together with the relative locations of the optical fiber and weight sensor, is illustrated in Figure~\ref{fig:load_diagram}. The subsystem is capable of applying compressive loads of up to 10~kg.

\subsubsection{Detection Systems}
\label{sec:detection}
\begin{figure}
\centering
\begin{subfigure}{0.25\textwidth}
    \includegraphics[width=\textwidth]{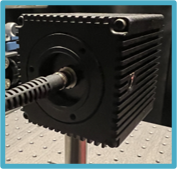}
    \caption{}
    \label{fig:detector}
\end{subfigure}
~
\begin{subfigure}{0.327\textwidth}
    \includegraphics[width=\textwidth]{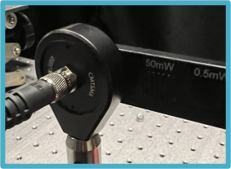}
    \caption{}
    \label{fig:output_photodiode}
\end{subfigure}

\caption{Detection systems used for FRD and relative throughput measurements. (a) Far-field image acquisition system, in which the optical fiber under test is coupled directly to the C-mount interface of the scientific CMOS detector. (b) Output optical-power measurement system used to measure the optical power emitted from the optical fiber for relative throughput characterization.}
\label{fig:detection_system}
\end{figure}

Figure~\ref{fig:detector} shows the far-field intensity-pattern acquisition setup used to measure FRD. The optical fiber under test is coupled to a C-mount fiber adapter plate, which is attached directly to the C-mount interface of the Tucsen Dhyana 401D scientific CMOS detector. This configuration ensures that the fiber output is directed onto the detector at normal incidence ($0^{\circ}$). For all measurements, the detector was operated in 16-bit mode with the gain set to HDR.

Figure~\ref{fig:output_photodiode} shows the output optical-power measurement setup. A Thorlabs S130VC photodiode power sensor is calibrated to the wavelength of the light source prior to each set of measurements and is used to measure the optical power emitted from the output end of the optical fiber. As discussed in Section~\ref{sec:input}, simultaneous measurements of the input and output optical powers are used to calculate the relative throughput of the optical fiber.

\subsection{Software Architecture}
A custom Python software framework was developed to coordinate the hardware components, automate data acquisition, and perform data analysis. The measurement workflow begins by reading user-defined test conditions from a graphical user interface. These conditions include the selected illumination wavelength, input angles, twist angles, bend radii, applied loads, and the number of exposures to acquire for each measurement configuration.

Based on the selected test conditions, the motorized subsystems are commanded to the required positions corresponding to the desired input angle, bend radius, twist angle, and applied load. Once the system reaches the requested configuration, data acquisition begins automatically.

For FRD measurements, a bias frame is acquired at the start of the measurement sequence. At each measurement configuration, an initial light frame is captured using an automated exposure-control algorithm to determine an appropriate detector exposure time. The remaining light frames are then acquired using the same exposure time until the specified number of exposures has been collected. A dark frame is subsequently captured at the same exposure time with the shutter closed. The bias, light, and dark frames, together with the corresponding measurement metadata, are stored as FITS files for subsequent analysis. The motorized subsystems are then repositioned to the next measurement configuration, and the acquisition process is repeated.

The automated exposure-control algorithm uses a binary-search strategy to determine an appropriate detector exposure time. The algorithm searches for an exposure time that produces a maximum pixel intensity between 80\% and 95\% of the 16-bit full-scale value of 65535 ADU. This prevents image saturation while maintaining sufficient signal for accurate FRD measurements.

For relative throughput measurements, the input and output optical powers are measured simultaneously using the photodiode sensors described in Sections~\ref{sec:input} and \ref{sec:detection}. The resulting optical-power data are stored with the corresponding measurement metadata for subsequent analysis.

\begin{figure}[ht]
\centering
\includegraphics[width=0.8\textwidth]{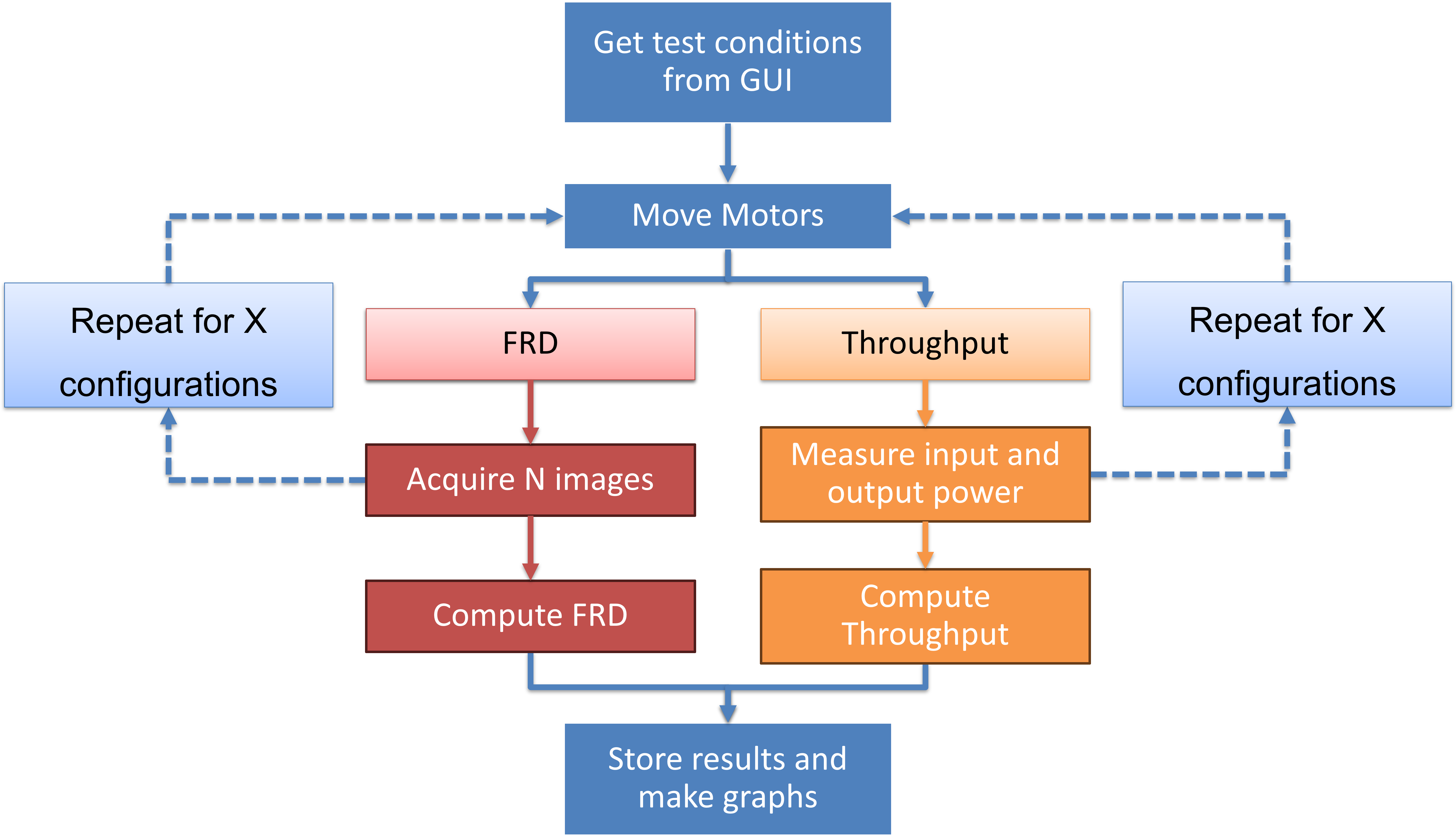}
\caption{Flowchart summarizing the software architecture of the automated FRD and relative throughput characterization system.}
\label{fig:software_flow_chart}
\end{figure}

Upon completion of the measurements, the acquired images and optical-power data are analyzed and used to generate summary plots. A detailed description of the FRD and relative throughput analysis procedures is provided in Section~\ref{sec:data_analysis}. The overall software architecture is summarized in Figure~\ref{fig:software_flow_chart}.

This automated workflow enables unattended characterization campaigns involving hundreds of measurement configurations, substantially reducing the time required for optical fiber characterization.

\subsection{Data Analysis}
\label{sec:data_analysis}

\subsubsection{FRD Measurement}
The analysis procedure was adapted from the collimated beam (ring) test method used for astronomical fiber FRD characterization \cite{Carrasco1994,Monty2018}. In this approach, a collimated beam is injected into the fiber at a known angle, and the resulting far-field ring diameter and radial thickness are used to quantify FRD.

\begin{figure}[htb]
\centering

\begin{subfigure}{0.32\textwidth}
    \includegraphics[width=\linewidth]{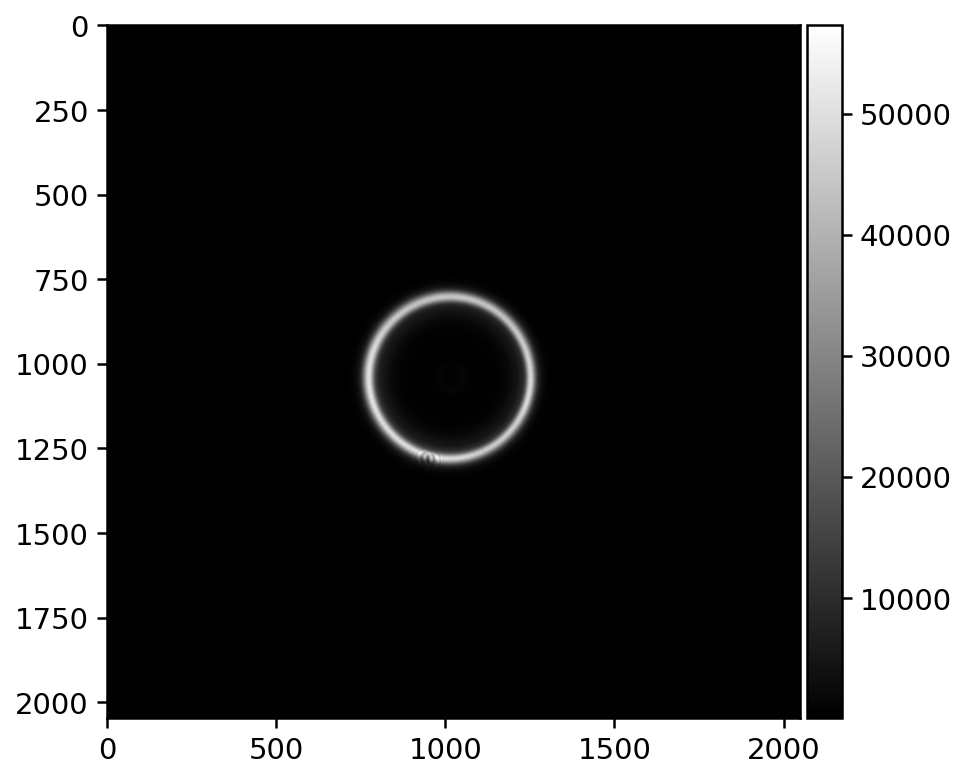}
    \caption{Light frame.}
    \label{fig:light_frame}
\end{subfigure}%
~
\begin{subfigure}{0.32\textwidth}
    \includegraphics[width=\linewidth]{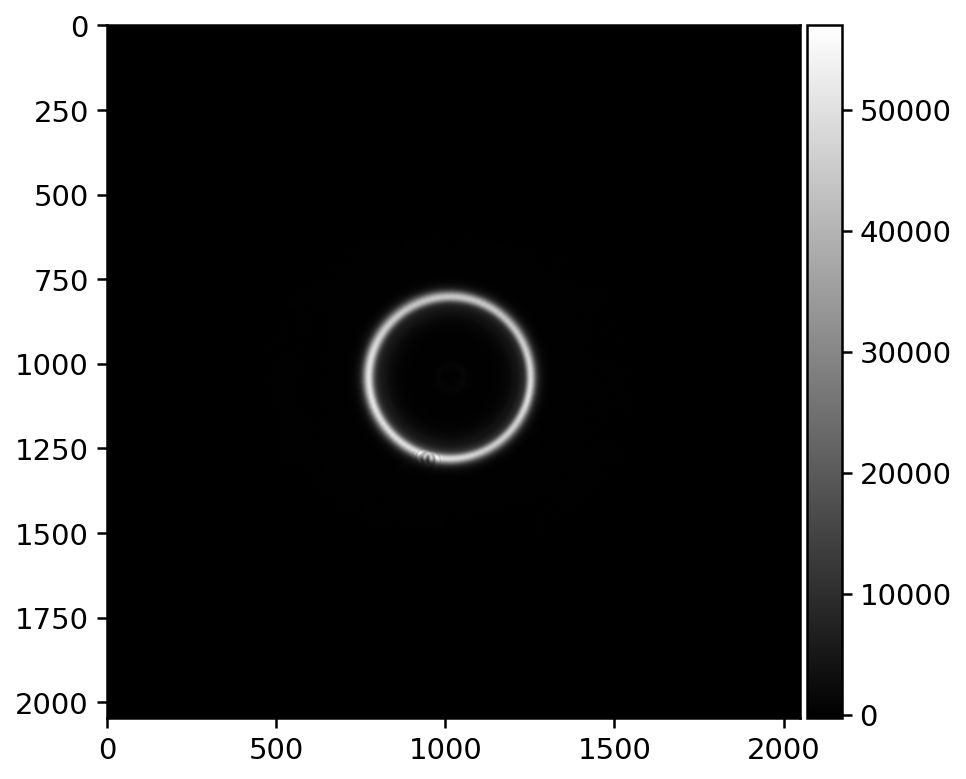}
    \caption{Dark- and bias-corrected image.}
    \label{fig:corrected_frame}
\end{subfigure}
\\
\begin{subfigure}{0.32\textwidth}
    \includegraphics[width=\linewidth]{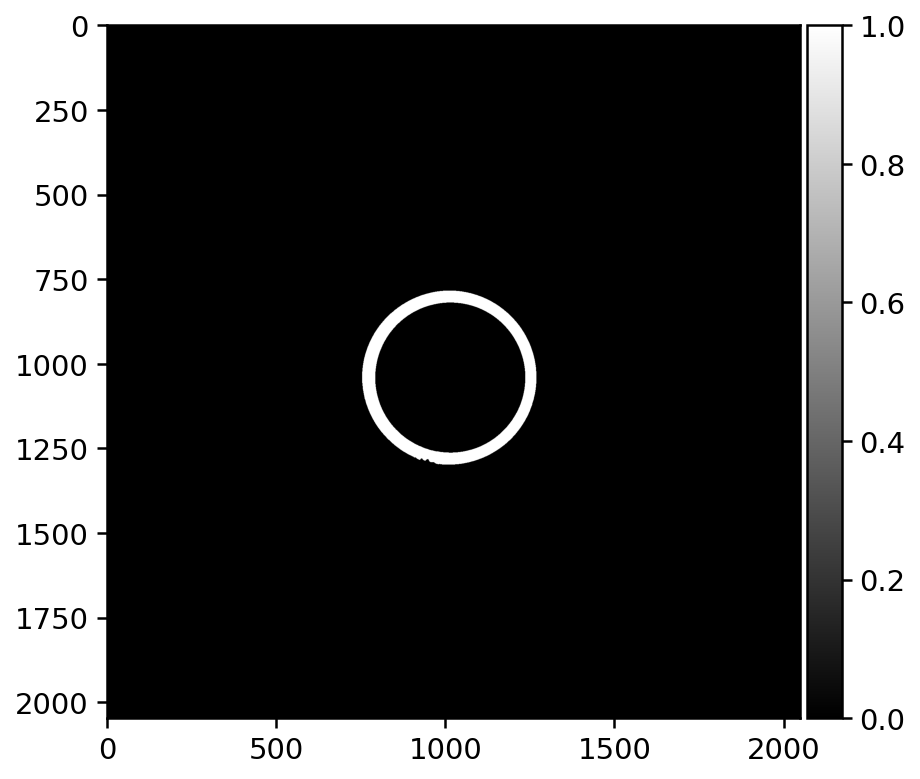}
    \caption{Binary mask generated using a threshold of 17097~ADU.}
    \label{fig:thresholded_frame}
\end{subfigure}%
~
\begin{subfigure}{0.32\textwidth}
    \includegraphics[width=\linewidth]{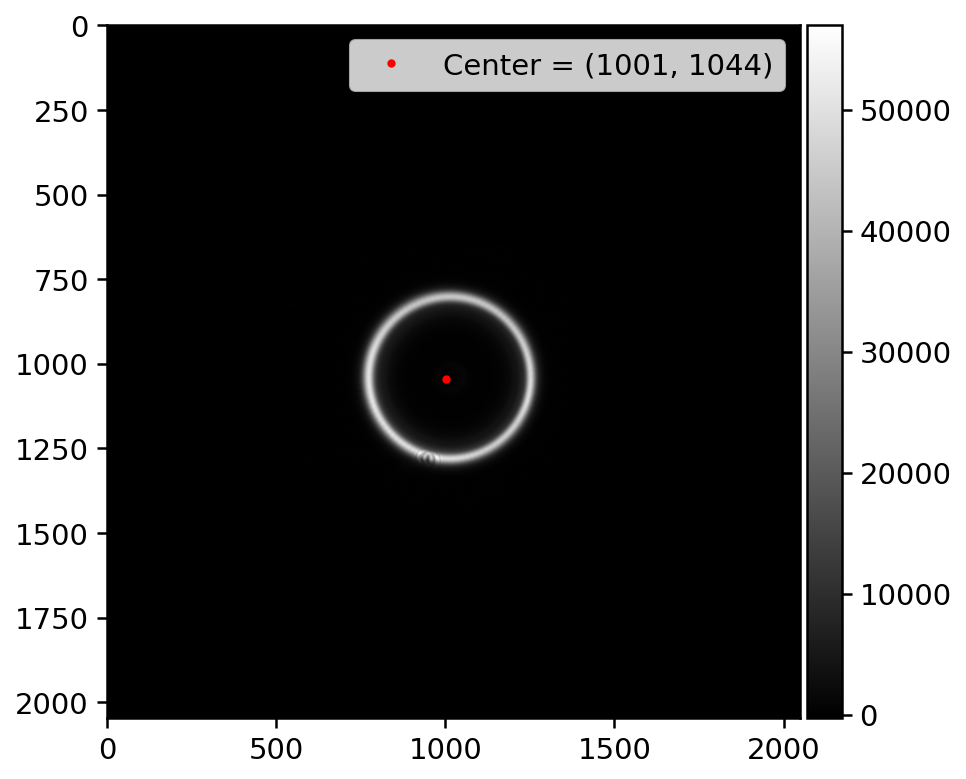}
    \caption{Ring center highlighted in red at $(1001, 1044)$.}
    \label{fig:center_frame}
\end{subfigure}
\\
\begin{subfigure}{0.64\textwidth}
    \includegraphics[width=\linewidth]{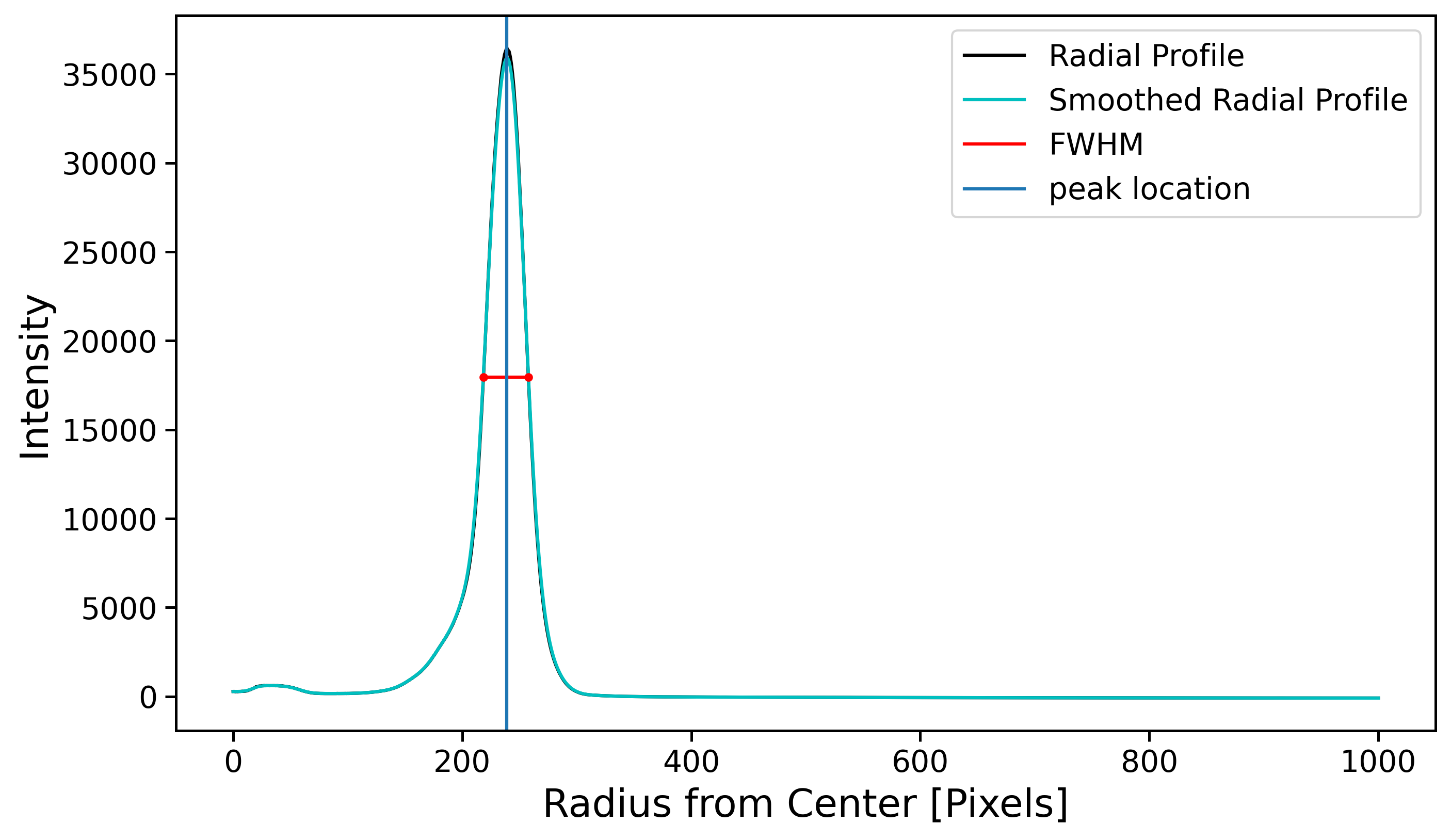}
    \caption{Radial profile, smoothed radial profile, FWHM, and peak location of the corrected image.}
    \label{fig:radial_profile}
\end{subfigure}

\caption{Example of the image-analysis procedure used to calculate focal ratio degradation (FRD) from a far-field intensity pattern.}
\label{fig:setup}
\end{figure}

Prior to analysis, each light frame was corrected for detector bias and background light. A bias frame was subtracted to remove the detector bias offset, and a bias-corrected dark/background estimate acquired at the same exposure time as the light frame was subtracted to remove exposure-dependent background contributions. An example of this correction is shown in Figures~\ref{fig:light_frame} and~\ref{fig:corrected_frame}.

The center of the far-field intensity pattern was then determined from the corrected image. The image was converted into a binary mask by assigning pixels with intensities greater than 30\% of the maximum pixel intensity a value of 1, while all remaining pixels were assigned a value of 0, as shown in Figure~\ref{fig:thresholded_frame}. This procedure isolates the ring-shaped far-field intensity pattern from the background. The center coordinates of the ring were calculated as the mean $x$- and $y$-coordinates of all pixels assigned a value of 1 in the binary mask. An example of the resulting center determination is shown in Figure~\ref{fig:center_frame}.

Using the calculated center coordinates, a radial intensity profile was generated by averaging the pixel intensities over concentric annuli centered on the far-field pattern. Successive annuli were separated by one pixel in radius. For each radius, the mean and standard deviation of the pixel intensities were calculated, yielding both the radial profile and an empirical estimate of the scatter associated with each radial intensity value.

The resulting radial intensity profile contained an approximately Gaussian peak corresponding to the ring-shaped far-field pattern. To reduce the influence of pixel-scale fluctuations, the profile was smoothed prior to analysis. The peak location was identified and used to determine the radius of the far-field ring. The full width at half maximum (FWHM) of the radial profile was then determined by identifying the two radial positions corresponding to half of the peak intensity and calculating their separation. An example radial profile is shown in Figure~\ref{fig:radial_profile}.

Uncertainties in the FWHM were estimated using a Monte Carlo approach. The radial intensity profile was repeatedly resampled based on the measured scatter associated with each radial intensity value, and the FWHM was recalculated for each realization. The resulting distribution of FWHM values was used to determine the FWHM uncertainty.

The measured FWHM was converted to the corresponding $1/e^2$ beam diameter and subsequently used with Equation~\ref{eq:FRD_equation} to calculate the FRD. Standard uncertainty propagation techniques were then applied to determine the uncertainty in the final FRD measurement.

\subsubsection{FRD Measurement Validation}
\begin{figure}[htb]
\centering
\includegraphics[width=0.49\textwidth]{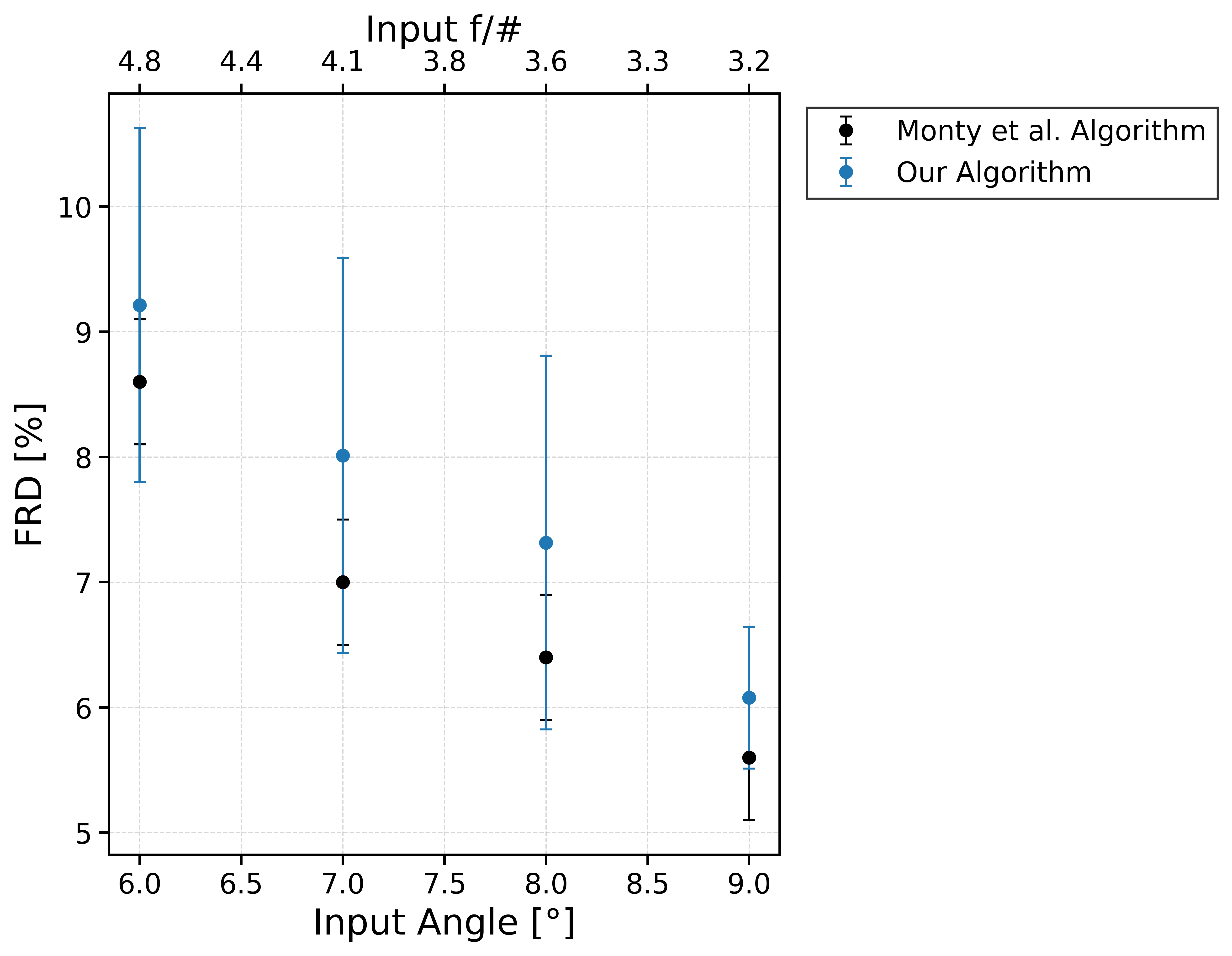}
\caption{Comparison of FRD values measured using the analysis algorithm developed in this work and the algorithm presented by Monty et al.~\cite{Monty2018}. Agreement between the two methods within the measurement uncertainties supports the accuracy of the FRD analysis pipeline implemented in this system.}
\label{fig:validation}
\end{figure}

To validate the FRD analysis procedure, the algorithm was applied to the sample far-field intensity patterns provided by Monty et al.~\cite{Monty2018}, for which reference FRD measurements are available. The resulting FRD values are compared with the values reported by Monty et al. in Figure~\ref{fig:validation}.

The FRD measurements obtained using the present analysis procedure are consistent with those reported by Monty et al.~\cite{Monty2018}, with agreement observed within the measurement uncertainties. Since Monty et al. demonstrated that their automated FRD measurements are in close agreement with manually determined FRD values, this comparison supports the accuracy of the FRD analysis procedure implemented in this work.

\subsubsection{Relative Throughput Measurement}
The input and output optical powers were measured simultaneously using the photodiode sensors described in Sections~\ref{sec:input} and \ref{sec:detection}. Simultaneous measurements were made possible by the nominal 50:50 beam splitter, which directed a portion of the incident light to the input photodiode while the remaining light was directed toward the optical fiber. For each relative throughput measurement, the input and output optical powers were recorded over a period of 1~s. The mean optical power during this interval was taken as the measured value, while the standard deviation of the measurements was used to estimate the associated uncertainty. 

The relative throughput of the optical fiber was then calculated using Equation~\ref{eq:throughput_equation}. Uncertainties in the relative throughput measurements were determined through standard uncertainty propagation of the input and output optical-power uncertainties.

\section{SYSTEM VALIDATION}

\subsection{Methods}
System validation was performed using a Thorlabs M14L02 multimode optical fiber with a 50~$\mu$m core diameter, a 125~$\mu$m cladding diameter, and a 3~mm protective jacket diameter \cite{Thorlabs_M14L02}.

To evaluate the performance of the automated characterization system, four experimental studies were conducted in which both FRD and relative throughput were measured:
\begin{enumerate}
\item dependence on input angle, with FRD measured for input angles satisfying $3^{\circ} \leq |\theta| \leq 10^{\circ}$ and relative throughput measured for input angles between $-10^{\circ}$ and $10^{\circ}$;
\item dependence on externally applied compressive load for loads between 0~g and 10,000~g;
\item dependence on fiber twist angle for angles between $-360^{\circ}$ and $360^{\circ}$;
\item dependence on fiber bend radius for radii between 5~mm and 35~mm.
\end{enumerate}

Each study was repeated at three illumination wavelengths:
\begin{enumerate}
\item 484.9~nm (25~nm bandwidth);
\item 594.2~nm (14~nm bandwidth);
\item 697.6~nm (18~nm bandwidth).
\end{enumerate}

For each FRD measurement configuration, five light-frame exposures were acquired. As described in Section~\ref{sec:input}, relative throughput measurements were obtained by simultaneously recording the input and output optical powers over a 1~s interval. The reported FRD for each configuration was calculated as the mean of the five measurements, with the associated uncertainty taken as the standard deviation of those measurements. The reported relative throughput was calculated from the mean input and output optical powers measured during the sampling interval, with the associated uncertainty determined through standard uncertainty propagation of the measured power fluctuations.

Because the optical fiber does not couple all of the incident light into its core, the optical power measured by the input photodiode is not equal to the optical power entering the fiber. Consequently, the throughput calculated using Equation~\ref{eq:throughput_equation} does not represent the absolute transmission efficiency of the optical fiber. Instead, it provides a relative throughput measurement. For the bend-radius, twist-angle, and externally applied-load studies, the fiber coupling geometry was held constant so that changes in the measured relative throughput could be attributed to the applied stressor. For the input-angle study, the injection geometry was intentionally varied, and the measured relative throughput therefore reflects changes in both coupling efficiency and optical transmission through the fiber.

The validation campaign consisted of 153 unique FRD measurement configurations, resulting in a total of 765 FRD measurements. In addition, 168 unique relative throughput measurement configurations were investigated. This dataset demonstrates the capability of the automated characterization system to efficiently perform large-scale optical fiber characterization with minimal user intervention.

\subsection{Results}
\label{sec:results}

\subsubsection{Dependence on Input Angle}
\label{sec:input_angle_dependence}
\begin{figure}[htb]
\centering
\includegraphics[width=0.49\textwidth]{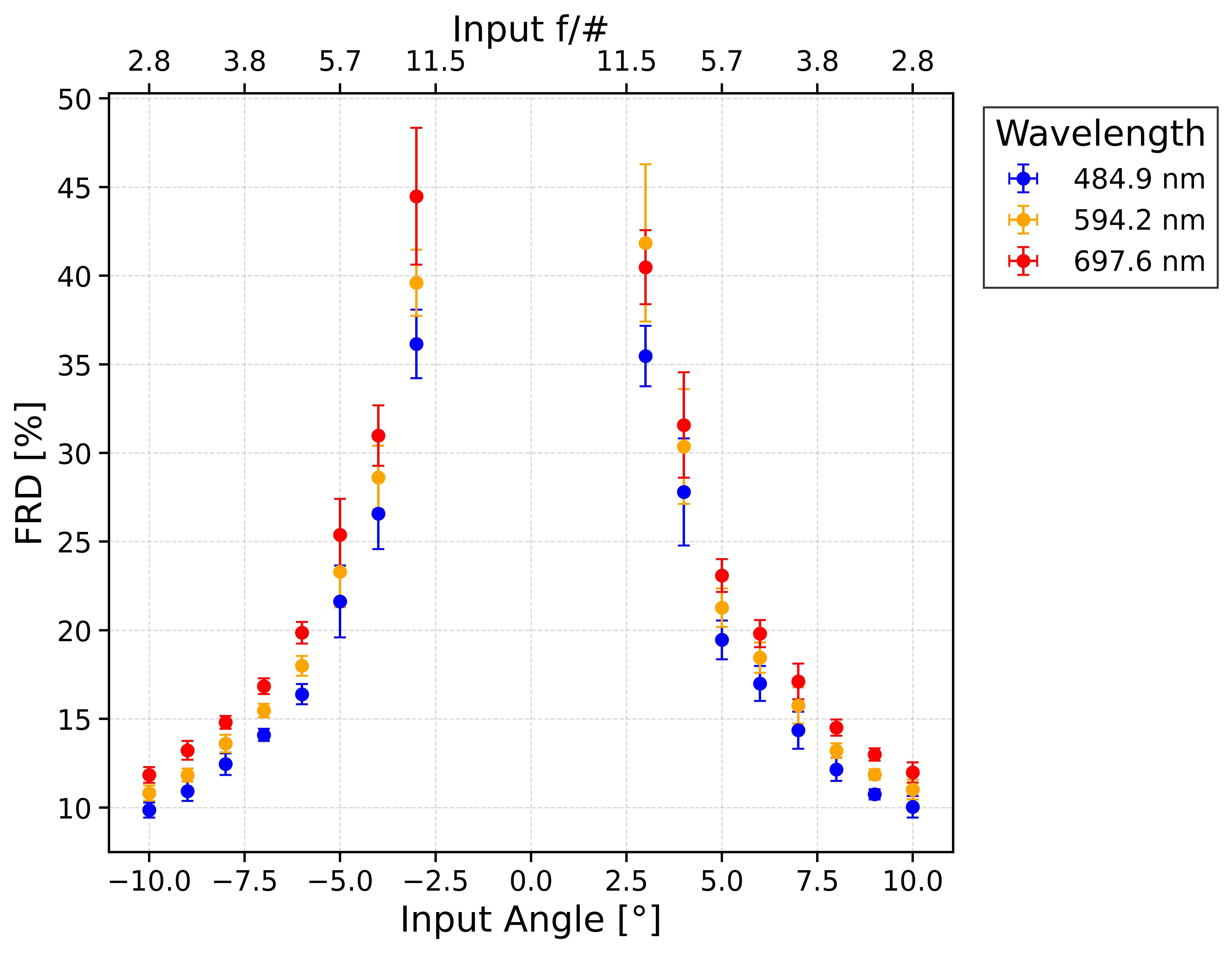}
\caption{Measured relationship between input angle and FRD. Measurements were obtained using light sources with wavelengths of 484.9~nm (25~nm bandwidth), 594.2~nm (14~nm bandwidth), and 697.6~nm (18~nm bandwidth). A twist angle of $0.00^{\circ}\pm0.09^{\circ}$, a bend radius of $36\pm1$ mm, and an applied load of 0 g were maintained throughout the experiment.}
\label{fig:angle_frd}
\end{figure}

As shown in Figure~\ref{fig:angle_frd}, the measured FRD exhibits a strong dependence on input angle, and therefore on input focal ratio. This behavior is consistent with the expected relationship between injection conditions and FRD. In addition, the measurements indicate that FRD is generally larger at longer wavelengths than at shorter wavelengths. These results demonstrate the importance of characterizing optical fibers under realistic injection conditions representative of those encountered in astronomical instrumentation.

\begin{figure}
\centering

\begin{subfigure}{0.49\textwidth}
    \includegraphics[width=\textwidth]{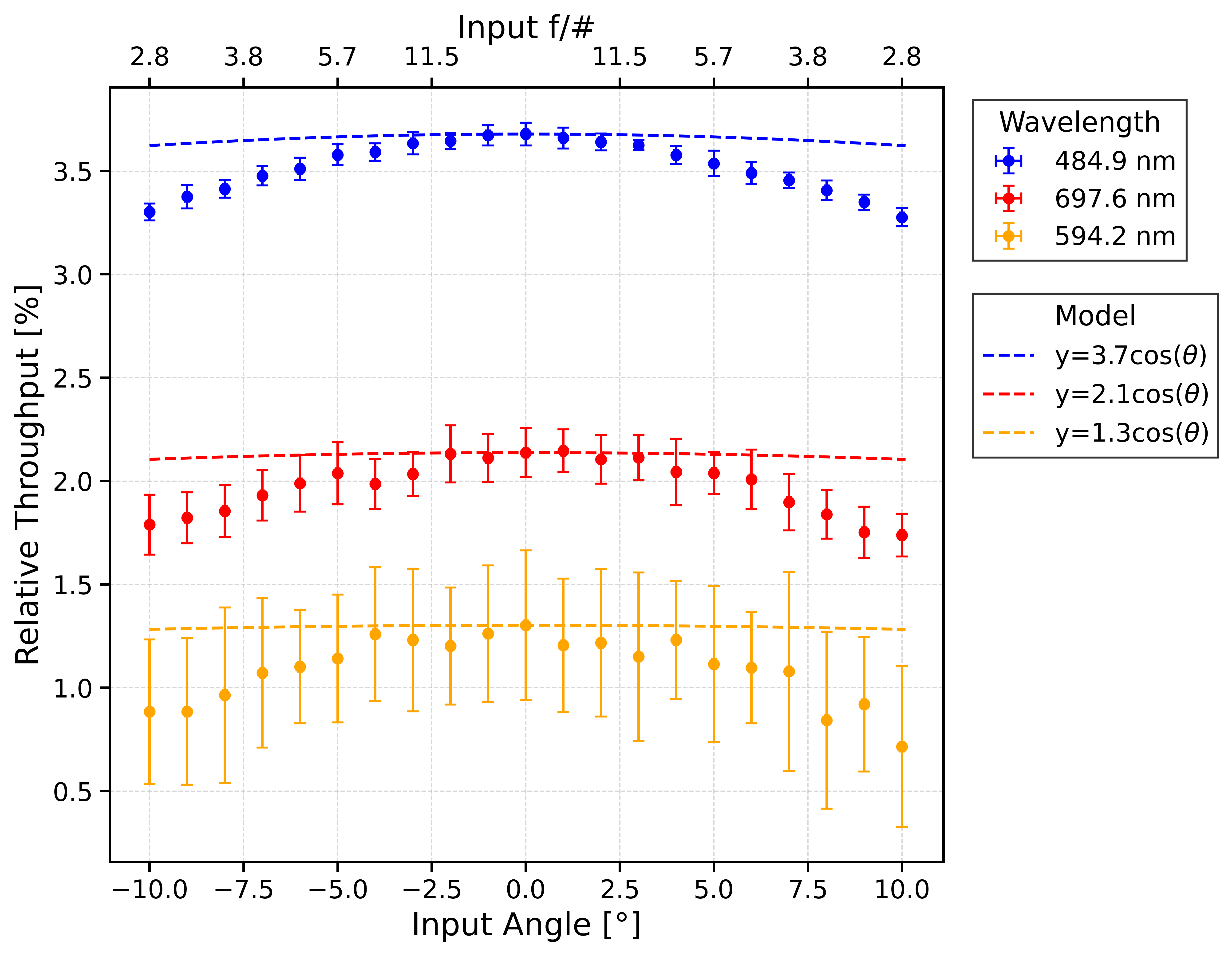}
    \caption{}
    \label{fig:angle_tp}
\end{subfigure}
~
\begin{subfigure}{0.49\textwidth}
    \includegraphics[width=\textwidth]{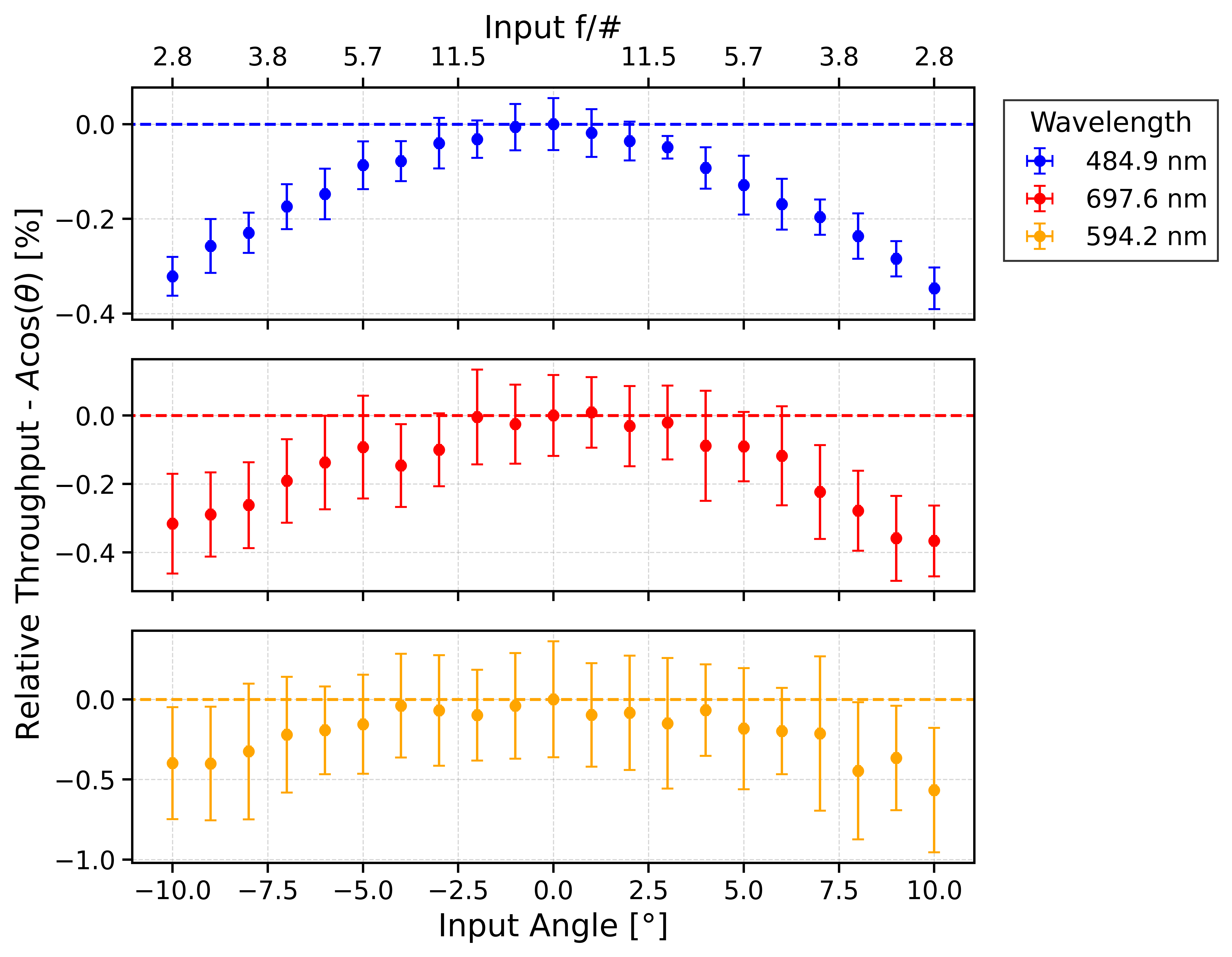}
    \caption{}
    \label{fig:angle_tp_residuals}
\end{subfigure}

\caption{Measured dependence of relative throughput on input angle. (a) Measured relative throughput as a function of input angle. The dotted curves show the projected-area model, $y=A\cos\theta$, where $A$ is the measured relative throughput at an input angle of $0^\circ$. (b) Residuals between the measured relative throughput and the projected-area model shown in (a). Measurements were obtained using light sources with wavelengths of 484.9~nm (25~nm bandwidth), 594.2~nm (14~nm bandwidth), and 697.6~nm (18~nm bandwidth). A twist angle of $0.00^{\circ}\pm0.09^{\circ}$, a bend radius of $36\pm1$ mm, and an applied load of 0 g were maintained throughout the experiment.}
\label{fig:angle_graphs}
\end{figure}

Figure~\ref{fig:angle_tp} shows the measured relative throughput as a function of input angle. For comparison, the dotted curves show the expected relative throughput assuming a simple projected-area model, $$y=A\cos\theta,$$ where $y$ represents the relative throughput and $A$ is the measured relative throughput at an input angle of $0^{\circ}$. Each model curve is shown in the same color as the corresponding measured dataset.

The residuals between the measured relative throughput and the projected-area model are shown in Figure~\ref{fig:angle_tp_residuals}. Although the measured relative throughput decreases with increasing $|\theta|$, the residuals indicate that the projected-area model does not fully reproduce the observed behavior within the experimental uncertainties. This result indicates that the reduction in relative throughput cannot be explained solely by the decrease in projected area available for coupling light into the optical fiber. Additional effects, such as changes in coupling efficiency, modal excitation, or light propagation within the fiber, may also contribute to the observed behavior.

The measured relative throughput at 484.9~nm (25~nm bandwidth) is consistently higher than that measured at 594.2~nm (14~nm bandwidth) and 697.6~nm (18~nm bandwidth), while the relative throughput at 697.6~nm (18~nm bandwidth) is consistently higher than at 594.2~nm (14~nm bandwidth). Because the reported values represent relative throughput rather than absolute fiber transmission efficiency, additional measurements would be required to determine whether this wavelength dependence arises from the fiber itself, wavelength-dependent coupling, or systematic effects in the measurement setup.

\subsubsection{Dependence on Twist Angle}
\begin{figure}[htb]
\centering

\begin{subfigure}{0.49\textwidth}
    \includegraphics[width=\textwidth]{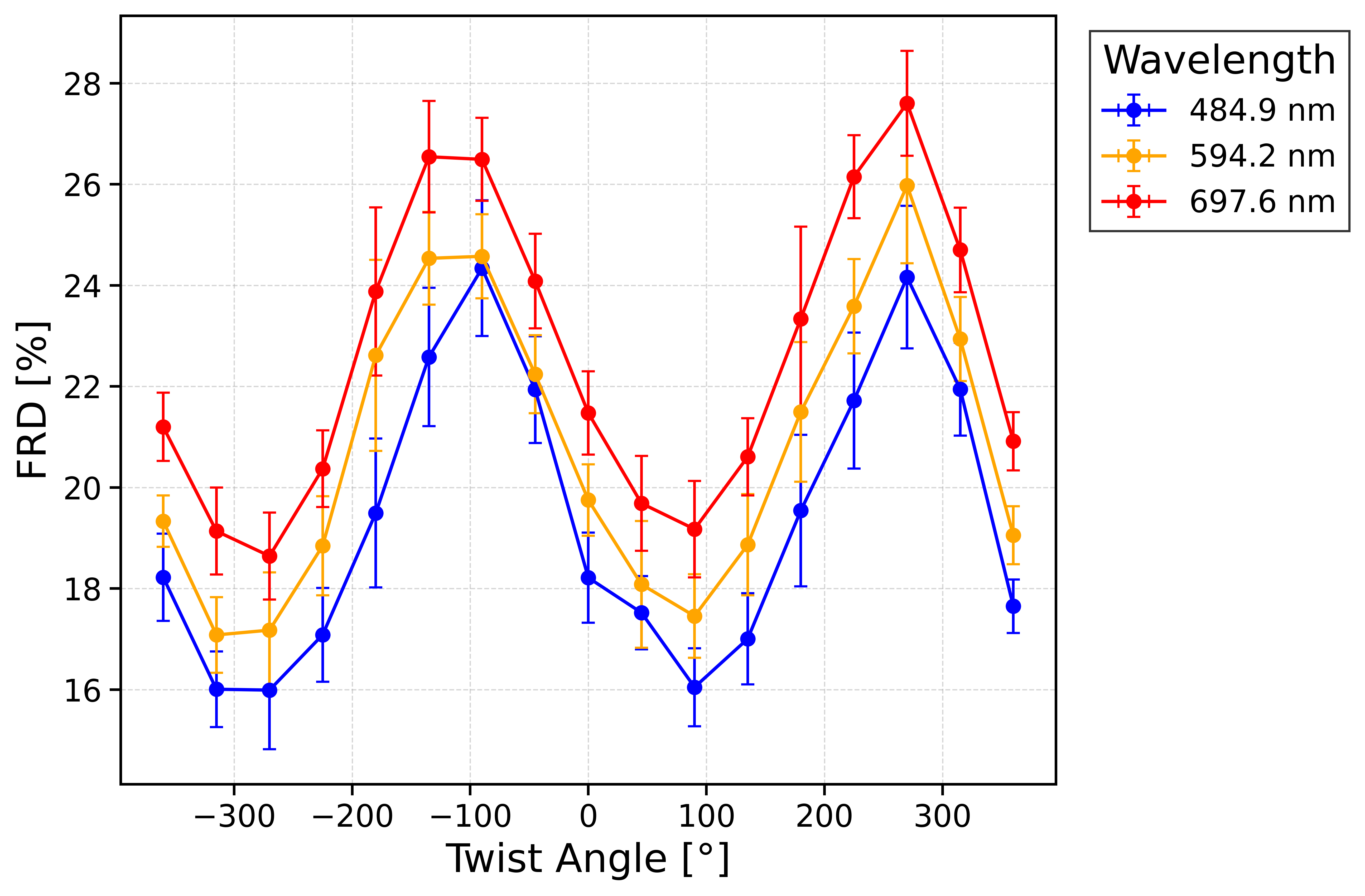}
    \caption{}
    \label{fig:twist_frd}
\end{subfigure}
~
\begin{subfigure}{0.49\textwidth}
\includegraphics[width=\textwidth]{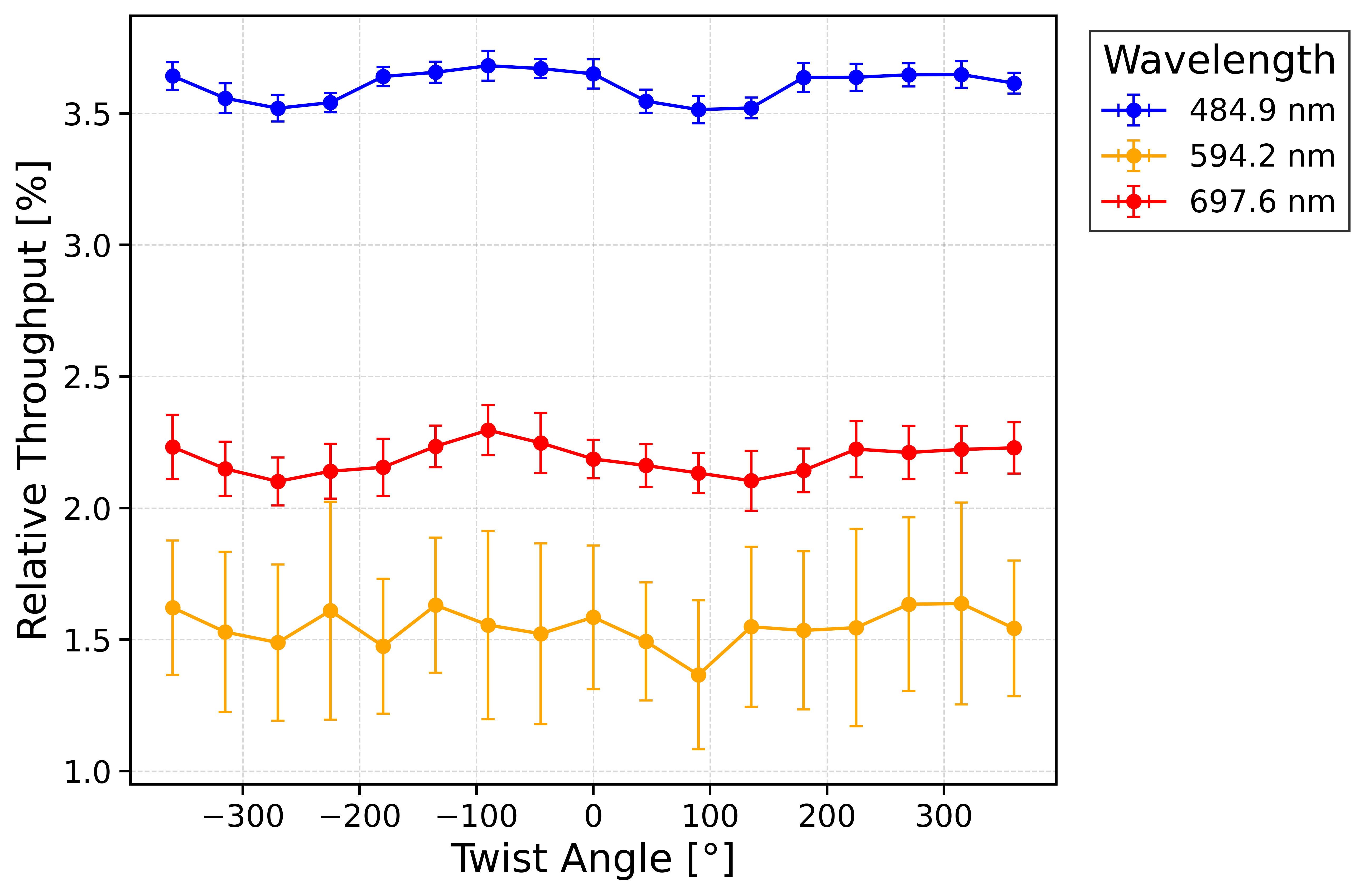}
    \caption{}
    \label{fig:twist_tp}
\end{subfigure}

\caption{Measured dependence of (a) FRD and (b) relative throughput on fiber twist angle. Measurements were obtained using light sources with wavelengths of 484.9~nm (25~nm bandwidth), 594.2~nm (14~nm bandwidth), and 697.6~nm (18~nm bandwidth). The input angle was maintained at $4.40^{\circ}\pm0.05^{\circ}$, while the bend radius and applied load were fixed at $36\pm1$ mm and 0 g, respectively.}
\label{fig:twist_graphs}
\end{figure}

As shown in Figure~\ref{fig:twist_frd}, the measured FRD exhibits a periodic dependence on fiber twist angle. The observed periodicity was reproducible across repeated measurements, suggesting that it is not attributable to measurement noise alone. The origin of this behavior was not investigated in detail; however, it may arise from asymmetries in the optical fiber or experimental setup that vary as the fiber is rotated about its optical axis. Despite this periodic variation, the overall magnitude of the FRD remained within a relatively narrow range over the investigated twist-angle range.

Figure~\ref{fig:twist_tp} shows that the measured relative throughput remained largely unchanged as the twist angle was varied. However, a weak periodic variation is observed in the measurements acquired at 484.9~nm (25~nm bandwidth), consistent with the periodic structure observed in the corresponding FRD measurements.

The wavelength dependence observed in both the FRD and relative throughput measurements is consistent with the trends discussed in Section~\ref{sec:input_angle_dependence}. 

\subsubsection{Dependence on Bend Radius}
\begin{figure}[htb]
\centering
\begin{subfigure}{0.49\textwidth}
    \includegraphics[width=\textwidth]{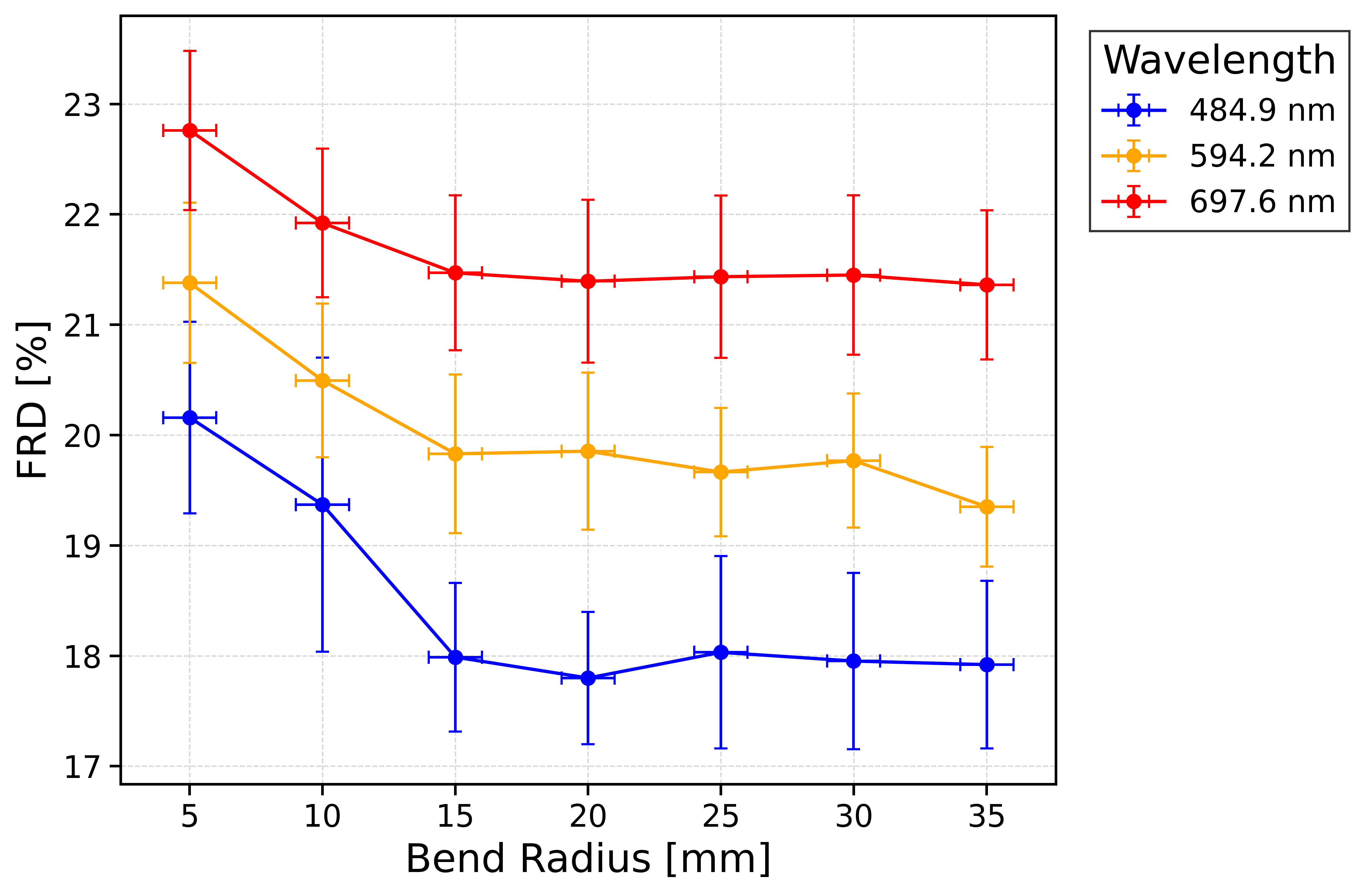}
    \caption{}
    \label{fig:bend_frd}
\end{subfigure}
~
\begin{subfigure}{0.49\textwidth}
    \includegraphics[width=\textwidth]{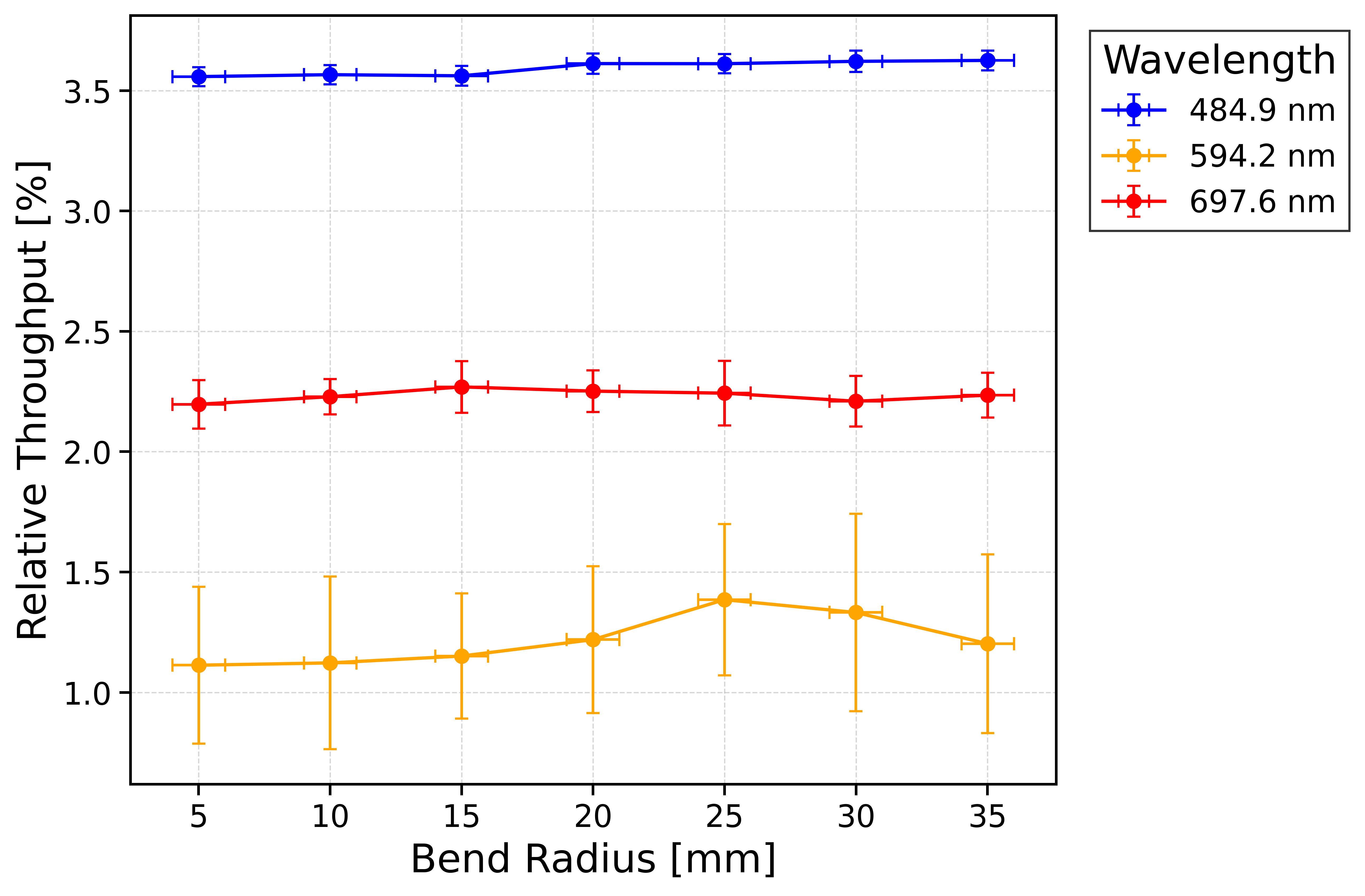}
    \caption{}
    \label{fig:bend_tp}
\end{subfigure}

\caption{Measured dependence of (a) FRD and (b) relative throughput on fiber bend radius. Measurements were obtained using light sources with wavelengths of 484.9~nm (25~nm bandwidth), 594.2~nm (14~nm bandwidth), and 697.6~nm (18~nm bandwidth). The input angle was maintained at $4.40^{\circ}\pm0.05^{\circ}$, while the twist angle and applied load were fixed at $0.00^{\circ}\pm0.09^{\circ}$ and 0 g, respectively.}
\label{fig:bend_graphs}
\end{figure}

Figure~\ref{fig:bend_frd} shows that reducing the bend radius resulted in a measurable increase in FRD. This effect became more pronounced at smaller bend radii, where bend-induced mode coupling is expected to increase. In contrast, little to no change in FRD was observed at larger bend radii, indicating that the optical fiber is relatively insensitive to bending over the larger bend radii investigated.

As shown in Figure~\ref{fig:bend_tp}, the relative throughput exhibited comparatively smaller variations with bend radius. A modest reduction in relative throughput was observed only at the smallest bend radii tested. The larger change observed in FRD compared with relative throughput suggests that bend-induced degradation is dominated primarily by modal redistribution rather than by significant optical power loss.

The wavelength dependence observed in both the FRD and relative throughput measurements is consistent with the trends discussed in Section~\ref{sec:input_angle_dependence}.

\subsubsection{Dependence on Applied Load}
\begin{figure}
\centering

\begin{subfigure}{0.49\textwidth}
    \includegraphics[width=\textwidth]{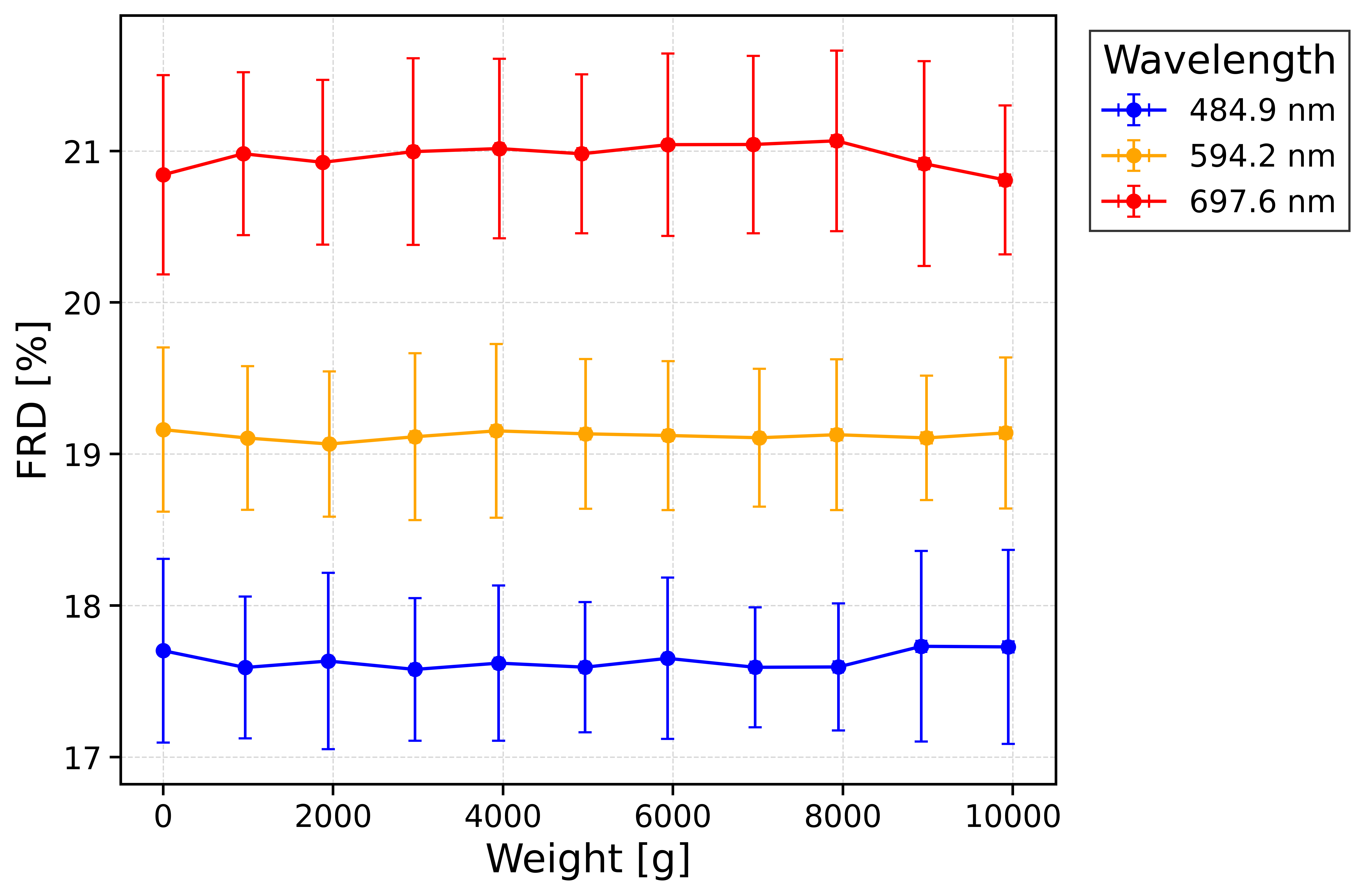}
    \caption{}
    \label{fig:weight_frd}
\end{subfigure}
~
\begin{subfigure}{0.49\textwidth}
    \includegraphics[width=\textwidth]{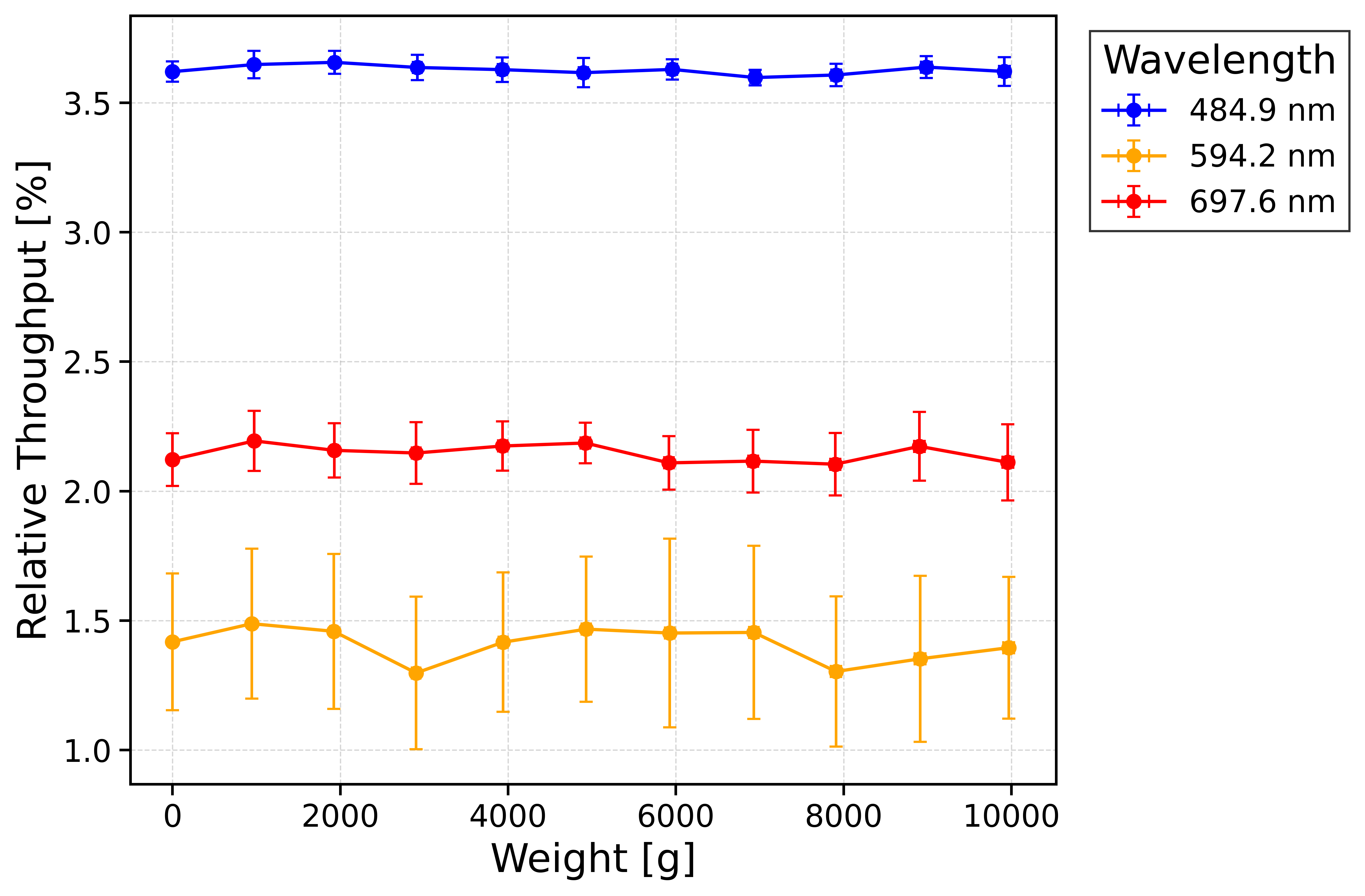}
    \caption{}
    \label{fig:weight_tp}
\end{subfigure}

\caption{Measured dependence of (a) FRD and (b) relative throughput on applied compressive load over a length of $100\pm1$ mm. Measurements were obtained using light sources with wavelengths of 484.9~nm (25~nm bandwidth), 594.2~nm (14~nm bandwidth), and 697.6~nm (18~nm bandwidth). The input angle was maintained at $4.40^{\circ}\pm0.05^{\circ}$, while the twist angle and bend radius were fixed at $0.00^{\circ}\pm0.09^{\circ}$ and $36\pm1$ mm, respectively.}
\label{fig:weight_graphs}
\end{figure}

Figures~\ref{fig:weight_frd} and~\ref{fig:weight_tp} show that both FRD and relative throughput remained effectively constant over the investigated load range. Within the measurement uncertainties, no measurable dependence of either FRD or relative throughput on externally applied compressive load was observed. These results suggest that moderate localized compressive loading is not a dominant contributor to FRD or relative throughput degradation for this fiber under the investigated conditions.

The wavelength dependence observed in both the FRD and relative throughput measurements is consistent with the trends discussed in Section~\ref{sec:input_angle_dependence}. 

\subsection{Discussion}

The measurements presented in Section~\ref{sec:results} support the broader view that FRD is a system-dependent property influenced by both intrinsic fiber characteristics and external integration conditions. Previous studies have identified several contributors to FRD, including microbending, end preparation, polishing quality, bonding length, angular misalignment, bending, torsion, and repeated mechanical flexure \cite{Ramsey1988,Oliveira2005,Poppett2010,AllingtonSmith2013,Eigenbrot2012,Murphy2013,Belland2020}. The present system was therefore designed to probe not only the nominal performance of an optical fiber, but also how that performance changes under controlled optical and mechanical perturbations. This distinction is important for astronomical instruments, where fibers are rarely operated as isolated components and are instead integrated into positioners, ferrules, bundles, routing systems, and spectrograph interfaces.

The strong dependence of FRD on input angle agrees with the expectation that injection conditions play a central role in determining the output angular distribution of light. This result is also consistent with instrument-focused studies showing that positioner operation, coupling geometry, and fiber-end positioning can complicate the interpretation of FRD measurements \cite{Belland2020,Yan2017}. The corresponding relative throughput measurements showed that the decrease in relative throughput with increasing input angle was not fully described by a simple projected-area model. This indicates that the observed relative throughput variation cannot be explained solely by the reduction in projected coupling area, and that additional effects, such as changes in coupling efficiency, modal excitation, or propagation within the fiber, may also contribute. Since the input-angle study intentionally changes the injection geometry, the measured relative throughput in this case should be interpreted as a combined measure of coupling and transmission rather than as an absolute fiber transmission efficiency.

The twist-angle measurements revealed a reproducible periodic dependence of FRD on fiber rotation, while the corresponding relative throughput remained nearly constant. Torsional effects are less commonly isolated in large-scale FRD studies than bending, polishing, or input-angle effects, making this observation particularly interesting. The reproducibility of the periodic structure suggests that the system is sensitive to small changes in the fiber or coupling geometry as the fiber is rotated. Possible explanations include azimuthal asymmetries in the fiber core, cladding, jacket, connector, stress distribution, or input coupling conditions. However, the physical origin of this behavior was not determined in this work. Additional tests using multiple fibers, reversed fiber orientations, independent alignment checks, and repeated mounting procedures would be required to determine whether the periodicity is intrinsic to the tested fiber or introduced by the experimental configuration.

The bend-radius measurements showed that decreasing the bend radius produced a measurable increase in FRD, while only modest reductions in relative throughput were observed. This behavior suggests that bending primarily redistributes optical power among guided modes, increasing the output angular spread without producing comparably large relative throughput losses over much of the investigated range. These results highlight the complementary nature of FRD and relative throughput measurements, as each provides different information regarding the effects of mechanical stress on optical fiber performance.

Within the investigated load range, externally applied compressive loading did not produce measurable changes in either FRD or relative throughput. This result should be interpreted within the specific loading geometry and fiber construction tested here. Previous studies have shown that stress introduced during mounting, polishing, bonding, or repeated motion can increase FRD, particularly near fiber ends or localized contact regions \cite{Oliveira2005,AllingtonSmith2013,Murphy2013}. The absence of a measurable load dependence in the present data therefore does not imply that compressive stress is generally unimportant. Rather, it suggests that the tested fiber, its protective jacket, and the applied-load configuration used in this study did not introduce sufficient microbending or deformation to measurably alter the output beam. More broadly, this result demonstrates that the characterization system can identify both significant performance changes and the absence of measurable degradation when a mechanical perturbation is insufficient to affect optical performance.

A consistent wavelength dependence was observed across the four experimental studies. In general, longer wavelengths exhibited larger measured FRD values, while the measured relative throughput at 484.9~nm remained higher than that at 594.2~nm and 697.6~nm. This wavelength dependence should be treated cautiously. Previous studies have reported a range of wavelength-dependent behaviors, with observed trends depending on fiber type, surface polish, input focal ratio, measurement method, and experimental configuration \cite{Murphy2008,Eigenbrot2012,Poppett2010}. Therefore, the wavelength trends observed here should be described as specific to the tested fiber and measurement setup, rather than as a general property of astronomical fibers. A broader wavelength sweep, absolute throughput calibration, and measurements of additional fiber types would be needed to determine whether the observed behavior is intrinsic to the fiber, stress-dependent, or associated with systematic effects in the measurement system.

Overall, these results demonstrate that the automated characterization system provides reliable and repeatable measurements of both FRD and relative throughput while requiring minimal user intervention. The system resolved strong effects, such as the dependence of FRD on input angle and bend radius, as well as subtler behaviors, such as the periodic twist-angle dependence. At the same time, it identified cases where no measurable degradation was observed, such as the applied-load study. The ability to automatically investigate multiple stressors, wavelengths, and measurement configurations makes the platform well suited for large-scale optical fiber characterization campaigns required by next-generation fiber-fed spectroscopic survey facilities.

\section{FUTURE WORK}
Future development of the characterization system will focus on expanding its measurement capabilities, increasing automation, and improving scalability for large optical fiber characterization campaigns. Planned investigations include:

\begin{enumerate} 
\item determining the origin of the observed twist-angle periodicity through measurements of additional fibers, reversed fiber orientations, and independent alignment configurations; 
\item developing an absolute throughput measurement capability through calibration of the fiber coupling efficiency and optical power reference path; 
\item characterizing fibers with different core diameters, numerical apertures, refractive-index profiles, jacket materials, and protective coatings; 
\item extending measurements over a broader spectral range to further investigate the observed wavelength dependence of FRD and relative throughput; 
\item adding automated tensile-stress and stretching characterization; 
\item performing long-term repeatability, stability, and environmental-sensitivity studies; 
\item extending the system to fiber bundles, multi-fiber assemblies, and fibers from different manufacturers and fabrication processes. 
\end{enumerate}

Future work will also focus on increasing measurement throughput through software optimization, improved hardware coordination, and potential hardware parallelization. Integration with database-driven analysis pipelines will enable automated storage, retrieval, comparison, and quality assessment of characterization data collected from large fiber populations. Collectively, these developments will support efficient characterization of the thousands of optical fibers required by next-generation fiber-fed spectroscopic survey facilities.

\section{CONCLUSION}
An automated multi-stressor optical fiber characterization platform capable of measuring both focal ratio degradation (FRD) and relative throughput under controlled optical and mechanical stress conditions has been developed. The system extends previous automated FRD characterization platforms by integrating wavelength selection, input-angle control, bend-radius control, twist-angle control, and externally applied compressive loading within a unified measurement framework. A custom software architecture coordinates hardware control, data acquisition, and analysis, enabling large characterization campaigns to be performed with minimal user intervention.

Validation measurements performed on a multimode optical fiber demonstrated that the platform can resolve both pronounced and subtle stress-dependent variations in fiber performance. The results confirmed the expected dependence of FRD on input angle and bend radius, while showing comparatively weak sensitivity to externally applied compressive loading under the investigated conditions. A reproducible periodic dependence of FRD on fiber twist angle was also observed, motivating further investigation using additional fibers and independent alignment checks. Simultaneous relative throughput measurements provided complementary information regarding the influence of each stressor on optical transmission and showed that relative throughput variations with incident angle cannot be explained solely by a simple projected-area model.

The automated workflow and scalability of the platform make it well suited for characterizing large populations of optical fibers. As upcoming spectroscopic survey facilities such as Spec-S5 and WST will require the characterization of tens of thousands of optical fibers and fiber-positioner assemblies, the system described in this work provides a practical framework for efficient, repeatable, and high-throughput optical fiber testing.

\acknowledgments

The authors acknowledge support from the David A. Dunlap Department of Astronomy and Dunlap Institute for Astronomy and Astrophysics at the University of Toronto. The Dunlap Institute is funded through an endowment established by the David Dunlap family and the University of Toronto.
This project was partially funded by the Canada Foundation for Innovation and Ontario Research Fund.

Generative artificial intelligence tools (ChatGPT, OpenAI) were used solely for language editing and grammar improvement during manuscript preparation. The authors reviewed and verified all content and assume full responsibility for the accuracy of the manuscript.

\bibliography{citations} 
\bibliographystyle{spiebib} 

\end{document}